\documentclass[useAMS,usenatbib]{mn2e}
\usepackage{epsfig}
\usepackage{longtable}

\def\gtsima{$\; \buildrel > \over \sim \;$}
\def\ltsima{$\; \buildrel < \over \sim \;$}
\def\gsim{\lower.5ex\hbox{\gtsima}}
\def\lsim{\lower.5ex\hbox{\ltsima}}
\def\kms{km s$^{-1}$}

\def\Lya{Ly$\alpha$}
\def\Lyb{Ly$\beta$}

\newcommand{\CIV}{\mbox{C\,{\sc iv}}}

\newcommand{\CII}{\mbox{C\,{\sc ii}}}

\newcommand{\OVI}{\mbox{O\,{\sc vi}}}
\newcommand{\OIV}{\mbox{O\,{\sc iv}}}

\newcommand{\MgII}{\mbox{Mg\,{\sc ii}}}

\newcommand{\SiIV}{\mbox{Si\,{\sc iv}}}
\newcommand{\SiIII}{\mbox{Si\,{\sc iii}}}
\newcommand{\SiII}{\mbox{Si\,{\sc ii}}}

\newcommand{\FeII}{\mbox{Fe\,{\sc ii}}}

\newcommand{\HI}{\mbox{H\,{\sc i}}}

\newcommand{\cloudy}{\mbox{\sc cloudy}}

\title[Metals in the $z\sim3$ intergalactic medium]{Metals in the $z\sim3$ intergalactic medium: results from an
  ultra-high signal-to-noise ratio UVES quasar
  spectrum}
\author[V. D'Odorico, et al.]{V. D'Odorico$^{1}$\thanks{E-mail:
dodorico@oats.inaf.it}, S. Cristiani$^{1,2}$, E. Pomante$^{1}$, R. F. Carswell$^3$, M. Viel$^{1,2}$, \and 
  P. Barai$^{1,4}$,  G. D. Becker$^{5}$, F. Calura$^{6}$,
  G. Cupani$^{1}$, F. Fontanot$^{1}$, 
   \and  M. G. Haehnelt$^{3,7}$, T-S. Kim$^{1}$, J. Miralda-Escud\'e$^{8,9}$,
   A. Rorai$^{3,7}$, E. Tescari$^{10,11}$, \and E. Vanzella$^6$  \\
$^1$ INAF- Osservatorio Astronomico di Trieste, Via Tiepolo 11, I-34143 Trieste, Italy \\
$^2$ INFN/National Institute of Nuclear Physics, via Valerio 2, I-34127 Trieste, Italy \\
$^3$Institute of Astronomy, Madingley Road, Cambridge CB3 0HA, UK \\
$^4$ Scuola Normale Superiore di Pisa, Piazza dei Cavalieri 7, I-56126 Pisa, Italy \\
$^5$ Department of Physics and Astronomy, University of California, Riverside, CA 92521, USA \\
$^6$ INAF - Osservatorio Astronomico di Bologna, Via Ranzani 1, I-40127 Bologna, Italy\\
$^7$ Kavli Institute of Cosmology, Madingley Road, Cambridge CB3 0HA, UK \\
$^8$ Instituci\'o Catalana de Recerca i Estudis Avan\c{c}ats, Passeig de Llu\'is Companys, 23, 08010, Barcelona, Catalonia, Spain \\
$^9$ Institut de Ci\`encies del Cosmos, Universitat de Barcelona, ICC-UB, Mart\'i i Franqu\`es 1, 08028, Barcelona, Catalonia, Spain \\
$^{10}$ School of Physics, The University of Melbourne, Parkville VIC
  3010, Australia \\
$^{11}$ ARC Centre of Excellence for All-sky Astrophysics (CAASTRO)\\
}
\begin{document}

\date{}

\pagerange{\pageref{firstpage}--\pageref{lastpage}} \pubyear{}

\maketitle

\label{firstpage}

\begin{abstract}
In this work, we investigate the abundance and distribution of metals
 in the intergalactic medium (IGM) at $\langle z \rangle \simeq 2.8$ through the analysis
 of an ultra-high signal-to-noise ratio UVES spectrum of the quasar
 HE0940-1050. In the \CIV\ forest, our deep spectrum  is sensitive at $3\ \sigma$ to lines with column density down to $\log N_{\rm CIV} \simeq 11.4$ and  in 60 per cent of the considered redshift range down to $\simeq11.1$. 
In our sample, all \HI\ lines with $\log N_{\rm HI} \ge 14.8$ show an
associated \CIV\ absorption. In the range $14.0 \le \log N_{\rm HI} <14.8$, 43 per cent of \HI\ lines has an associated \CIV\ absorption. At $\log N_{\rm HI} < 14.0$, the detection rates  drop to $<10$ per cent, possibly due to our sensitivity limits and not to an actual variation of the gas abundance properties.  
In the range $\log N_{\rm HI} \ge 14$, we observe a fraction of
\HI\ lines with detected \CIV\  a factor of 2 larger
than the fraction of \HI\ lines lying in the circum-galactic medium (CGM) of
relatively bright Lyman-break galaxies hosted by dark matter haloes
with $\langle M\rangle \sim10^{12}$ M$_{\sun}$. 
The comparison of our results with the output of a grid of
photoionization models and of two cosmological
simulations implies that the volume filling factor of the IGM  gas 
enriched to a metallicity $\log Z/Z_{\sun} \gsim -3$ should be of the
order of $\sim 10-13$ per cent.  
In conclusion, our  results  favour a scenario in which metals are found also outside the CGM of bright star-forming galaxies, possibly due to pollution by lower mass objects and/or to an early enrichment by the first sources.   
\end{abstract}


\section{Introduction}
The observed properties and evolution of galaxies in the Universe
require the existence of mechanisms regulating and quenching star
formation. Models and simulations which are not 
implementing these {\sl feedback} processes predict e.g. galaxies with far
larger stellar masses than are observed (e.g. Katz, Weinberg \& Hernquist
1996; Somerville \& Primack 1999; Cole et al. 2000; Springel \&
Hernquist 2003; Keres et al. 2009, and references therein). Stellar
winds, the explosion of supernovae and winds from active galactic nuclei (AGN) all
contribute to the heating and expulsion of gas from the disc and the
bulge of galaxies into the halo and possibly far away 
into the lower density gas. 
The signatures of strong winds are widely observed in local star
forming galaxies \citep[see e.g.][for a review]{winds05}. Their
presence at high redshift has been probed spectroscopically by the
observed offsets between the redshifts of the nebular emission lines
and those of the inter-stellar absorption lines in spectra of lensed
or extremely bright Lyman-break galaxies (LBGs, e.g. Pettini
et al. 2001; Shapley et al. 2003; Martin 2005; Rupke et al. 2005;
Tremonti et al. 2007; Weiner et al. 2009) and has been explored in cosmological hydrodynamical simulations \citep[e.g.][]{tescari11,barai13}.  
Recently, direct observational evidence for powerful, quasar-driven
outflows has emerged (e.g. Feruglio et al. 2010; Rupke \& Veilleux
2011; Sturm et al. 2011; Greene, Zakamska \& Smith 2012; Maiolino et al. 2012;
Cicone et al. 2014; Harrison et al. 2014). These outflows extend over
scales of several kpc from the nucleus and have been  
detected in both atomic and molecular gas. 

The indisputable evidence of the existence of galactic outflows
leaves open one fundamental question: how far is this gas travelling?   
Most of the times, the measured outflowing velocities are
large enough to allow the escape of the gas from the virial radius (e.g. Pettini
et al. 2001; Vanzella et al. 2010) and theoretical models predict that the expanding material may reach characteristic distances 
of a few hundred kpc and up to $\sim 1$ Mpc (e.g. Aguirre et al. 2001; Calura \& Matteucci 2006; Samui et al. 2008).


The presence of products of the
stellar nucleosynthesis outside galaxies has been confirmed by the
first high-resolution, high signal-to-noise ratio (SNR) optical spectra of
high redshift quasars used as background bright sources. In those
spectra, weak absorption features mainly due to the ionic doublet of
\CIV\ at $z_{\rm abs} \sim2-3$ ($\log N_{\rm CIV} \ge 12$) were detected at
the same redshift of about half of the \HI\ Lyman $\alpha$ (\Lya) lines with
column density $\log N_{\rm HI} \ge 14.5$ and of all \Lya\ lines with $\log
N_{\rm HI} \ge 15$ \citep{tytler95,cowie95,songaila_cowie96}.
In the same years, thanks to theoretical models and hydro-dynamical
simulations, the so called \Lya\ forest of \HI\ lines, observed in the
spectra of high-redshift quasars, 
was interpreted as due mainly to the density fluctuations in the
diffuse gas between galaxies: the inter-galactic medium
 (IGM, see e.g. Cen et al. 1994; Petitjean et al. 1995; or Rauch 1998; Weinberg et al. 1999, for
reviews). 
    
The presence of heavy elements in gas outside the virial radius of
galaxies confirmed observationally the effectiveness of outflows and
generated a new question: have the metals, that we observe associated with the
\Lya\ forest at $z\sim 2-3$, been ejected by the co-eval population of
star forming  galaxies  or are they due (also) to previous generations of
galaxies? 

Indeed, models of the \HI\ re-ionization process foresee that the same
sources which re-ionized the Universe were also able to pollute the
gas to a metallicity floor of the order $\log (Z/Z_{\sun}) \simeq
-3$ \citep[e.g.][]{madau01}. The volume filling factor of
the metals due to this pre-enrichment depends on the nature of the sources, with the
star-forming haloes with the smallest masses being the most efficient
pollutant of the IGM on large scales \citep[e.g.][]{Bertone2005}. 

To test  observationally this scenario, efforts have been concentrated
on two main approaches: the detection of metals in the outskirts of
galaxies and the investigation of the level of pollution in the low
density gas at overdensities of a few. 

Metals close to galaxies are generally searched for considering the
correlation between the metal absorptions observed along a quasar line of
sight and the galaxies at matching redshifts present in the field. All
the studies carried out up to now both at high \citep[e.g.][]{adelberger,steidel10,turner14} and low
redshift (e.g. Prochaska et al. 2011; Tumlinson et al. 2011;  Werk et al
2013; Bordoloi et al. 2014; Liang \& Chen 2014) agree on the
significant presence of metals in high and low ionization state
at impact parameters at least as large as $\approx 100-300$
kpc. However, this piece of evidence is not sufficient to state that
the observed metals were freshly ejected by the associated galaxies, since the same
correlation at $z\sim2-3$ could originate from the
accumulation of older metals in the knots of the cosmic web due to
gravitational collapse \citep[e.g.][]{porciani_madau}.
 Similar results for the size of the regions enriched by galaxies
  were found based on the clustering properties of metal absorbers (mainly 
\CIV\ lines) along \citep[e.g.][]{scannapieco06} and across
\citep[e.g.][]{martin10} the lines of sight to distant quasars.    

A few years after the discovery of the first metals polluting the IGM, 
the exploration of the low density regime was attempted using the stacking
method \citep{lu98} and the individual pixel optical depths (POD) of
\Lya\ and \CIV\ \citep{cowie_song98}. However, rigorous testing
of the analysis procedures by  \citet{ellison99} revealed that in
particular the stacking method suffered 
from severe limitations, leaving the question of whether or not the
low density regions of the IGM were enriched 
unanswered.

The next step to reach the metallicity level due to a possible pre-enrichment
was the observation  of an exceptionally high SNR spectrum of the
gravitationally lensed quasar B1422+231 ($z_{\rm em} =
3.623$) with the high resolution spectrograph HIRES
at the Keck telescope. \citet[][ESSP00 hereafter]{ellison00} showed
that the $5\ \sigma$ detection limit for 
\CIV\ outside the \Lya\ forest in this object is $\log N_{\rm CIV}
\simeq 11.6$ for a Doppler value $b=13$ \kms\ and the column density
distribution function corrected for completeness is fitted by the same
power law at least down to $\log N_{\rm CIV} = 11.75$, without showing a
downturn. The POD result is consistent with an almost constant
$\log ($\CIV/\HI$) \sim -2.6$ down to $\tau($\Lya$) \sim2-3$ which
implies a contribution of \CIV\ absorbers with column densities below
the detection limit.    

Further studies based on the POD technique have been carried out with
larger samples \citep{schaye03,aracil04}. In particular,
\citet[][S03 hereafter]{schaye03} based their study on the high SNR spectrum of
B1422+231 plus other 18 objects observed at lower SNR. They used
hydrodynamical simulations to convert the measurement of the relation
between the optical depths $\tau_{\rm CIV}$ and $\tau_{\rm HI}$ into a relation between
[C/H], overdensity\footnote{We define the overdensity as $(\delta + 1)
  \equiv \rho / \bar{\rho} $.} and redshift. They found that for a fixed
overdensity and redshift, the metallicity distribution is close to
lognormal with a large scatter (e.g. $\sigma([$C/H$]) \simeq 0.7$ at
$\delta =5$ and $z=3$); the median metallicity does not vary
significantly with redshift in the range $z=1.8-4.1$, while it
decreases with the overdensity:   

\begin{equation}
\label{eq:pod}
[\rm{C/H}] \simeq -3.47 + 0.65 (\log \delta -0.5) + 0.08(z-3)
\end{equation}

\noindent
and the relation is valid down to
approximately the mean density. This result, due to its statistical
nature, lacks the information on the distribution and covering factor
of metals which could only be derived by a direct detection of the very
weak metal lines and their association with the corresponding \Lya\ 
lines.


%
After the pioneer work by ESSP00,  the most ambitious surveys for weak
associated metal absorption features (based on Voigt profile fitting
of \CIV\ lines) have detected \CIV\ with $\log N_{\rm CIV} \gsim 12$
associated with \HI\ absorption with $\log N_{\rm HI} \gsim  14.5$
\citep[e.g.][]{dodorico10,BS15}, corresponding approximately  to
overdensities of a few at $z \sim 3$, probing only the densest few
per cent of the cosmic volume.  
At lower \HI\ column densities, the detection of associated \CIV\
absorption peters out \citep{cowie_song98}. 

The \OVI\ species is expected to be a more sensitive tracer of metals
in the low density IGM at $z\sim2-3$  \citep[e.g.][]{cen_chisari11},
but the relevant absorption lines occur in the same wavelength range
as the Lyman forest, which makes the detection of weak
\OVI\ absorption challenging  \citep[e.g.][]{bergeron02,carswell02,simcoe04}.
  
Moved by the motivation of directly investigating the metal enrichment of the IGM down to the mean density, we have repeated 
the enterprise of obtaining an ultra-high SNR quasar spectrum (dubbed
the {\it deep} spectrum) using UVES
at the ESO Very Large Telescope, 15 years after the observation
of the first and only extremely high SNR quasar spectrum of B1422+231. 

In this paper, we describe the observation campaign, the data
reduction and analysis of the UVES deep spectrum  of the quasar
HE0940-1050 ($z_{\rm em} = 3.0932$) in \S~2.  
Section~3 is dedicated to the statistics of the lines and the comparison with previous samples. In
\S~4, we present the \CIV\ detection rate and discuss the connection
with galaxies. Section~5 addresses the core topic, the metallicity of
the IGM. The final discussion and the conclusions are drawn in \S~6.          

The computation of the POD for \CIV\ and \OVI\ will be reported in a
further paper. The thermal properties of the IGM at $z\sim 3$ have
been studied with the \Lya\ forest of the deep spectrum in
\citet{rorai}. 

Throughout this paper, we assume $\Omega_{\rm m} = 0.26$,
$\Omega_{\Lambda} = 0.74$ and $h \equiv H_0/(100 {\rm km\ s}^{-1} {\rm
  Mpc}^{-1}) =0.72$.

\section{Data acquisition and treatment}
 We carefully chose the target for the deep spectrum to address two main
requirements: i) an emission redshift $z_{\rm em}\sim3$, representing
a trade-off between the need to probe the low-density medium (at
higher redshift the same column density corresponds to lower density)
and that of avoiding the 
excessive crowding in the \Lya\ forest (which increases with
redshift); ii) a line of sight free from Damped \Lya\ 
systems, whose presence would significantly decrease the portion of
\Lya\ forest available for our study. 

The selection was limited to the quasars observed for the UVES Large
Programme \citep[][from now on UVES LP]{bergeron04} because they were
all picked out to be bright, Damped-free targets and they have available good quality
spectra. We chose the quasar HE0940-1050 ($z_{\rm em} \simeq 3.0932$)
which is the brightest quasar of the sample fulfilling our
requirements. 

 Aiming at detecting at $3\ \sigma$ confidence level lines with column densities $\log N_{\rm CIV}
\simeq 11.5$ ($\log N_{\rm OVI} \simeq 12.2$)\footnote{If not
  otherwise specified, the sensitivities mentioned throughout the
  paper always refer to the strongest line of the doublet:
  \CIV\ $\lambda\,1548$ \AA\ and \OVI\ $\lambda\,1032$ \AA.}, we computed the required minimum SNR
 to be $\sim 205$ and 90 per resolution element in the \CIV\ and
 \OVI\ forests respectively, using the formula  \citep{herbertfort06}: 

\begin{equation}
\label{eq:ewlim}
{\rm SNR} \simeq \frac{3\, \lambda_{\rm r,X}}{c\, w_{\rm r,X}}\, \sqrt{4.24264\, b\, \Delta\,v}
\end{equation}

\noindent
where, 
$\lambda_{\rm r,X}$  and $w_{\rm r,X}$ are the rest frame wavelength
and  equivalent width of transition X, respectively, $c$ is the speed
of light, $b$ is the Doppler parameter of the lines (assumed to
be $ \sim 10$ \kms), and $ \Delta\,v$ is the velocity resolution element in \kms. 
The equivalent width of the line is then converted into a column
density assuming  the linear regime of the curve of growth:

\begin{equation}
\label{eq:cdlim}
N({\rm X}) = 1.13 \times 10^{20} w_{\rm r,X} / (f_{\rm X} \,
\lambda_{\rm r,X}^2) 
\end{equation} 

\noindent
where, $f_{\rm X}$ is the oscillator strength of transition X. 

\begin{table*}
\caption{Journal and settings of observations}
\begin{minipage}{100mm}
\label{tab_obs}
\begin{tabular}{l l  c c c c r}
\hline  
Prop. id & Date & Setup & Slit & CCD & No. of &  $T_{\rm exp}$ \\ 
& & (nm) & (as) & binning & frames & (s) \\
\hline
166.A-0106$^a$ & 2001 Feb  & 346-580 & 1.0 & 2x2& 4 & 14400 \\
                       & & 437-860 &  1.0 & 2x2& 4 & 14400 \\
079.B-0469$^b$ & 2007 Jun  & 390-564 & 1.0 & 2x2& 2 & 6000 \\
185.A-0745$^c$ & 2012 Mar-Apr  & 390-580 & 0.8 & 1x1 & 3 & 13200 \\
                           & & 390-585 & 0.8 & 1x1 & 10 & 47600 \\
092.A-0170$^d$ & 2013 Dec  & 390-564 & 1.0& 2x2& 1 & 4565 \\
                           & 2014 Jan-Mar  & 390-564 & 1.0 & 2x2 & 32 & 146080 \\                          
\hline
\end{tabular}
 $^a$\citet{bergeron04}; $^b$\citet{misawa}; $^c$\citet{molaro13}, $^d$our proposal 
\end{minipage}
\end{table*}

 Based on these computation, we proposed and obtained  43
hours of observations with UVES at the ESO VLT. 
The programme was carried out between 2013 December and 2014 March, adopting a 
slit of 1 arcsec, the CCD binned $2\times2$ and using the dichroic standard
setup with the blue arm centred at 390 nm and the red arm centred at
564 nm.  This configuration allows us to cover the two regions of
interest with one observation (bluewards of 360 nm the quasar flux
is almost completely absorbed by the break due to a Lyman limit
system). 

To create our final spectrum we used also all the other observations
of HE0940-1050 already present in the ESO UVES archive, for a total of
64.4 hours both in the region of the Lyman $\alpha$/Lyman $\beta$
forests and of the \CIV\ forest on the red side of the quasar
Lyman $\alpha$ emission.   The journal of observations and the specific
settings adopted by the different programmes are reported in
Table~\ref{tab_obs}.   

\subsection{Data reduction}

Our data plus all the data of the same object available in the ESO
UVES archive were reduced with the most recent version of the UVES
pipeline \citep[v. 5.4.3,][]{UVESpipe} in the context of the Reflex
workflow environment.  The set of spectra observed at higher resolution
(see Table~\ref{tab_obs}) was degraded to the resolution of all the
other spectra. All spectra were rebinned to 0.003 nm in the 346 and
390 setups, to 0.0035 nm in the 437 setup, and to 0.0045 nm in the 564, 580,
585 setups.  

A dedicated software developed by us \citep[see][]{Cupani15a,
  Cupani15b} for the future VLT spectrograph ESPRESSO \citep{pepe14} has been used in order to: 
\begin{itemize}
\item[i)] correct the blaze response and normalize each 1-D extracted,
  (vacuum heliocentric-)wavelength calibrated order of each
  spectrum. In this way all the spectral orders are brought to the
  same flux scale with the appropriate uncertainties for each pixel; 
\item[ii)] co-add in a single operation all the orders of all the
  spectra with a drizzling technique \citep[e.g.][]{Gonzaga12} to the
  final wavelength grid with a 2.5 \kms\ pixel size.  
\end{itemize}

 The flux in each velocity bin of the final spectrum is obtained from a
weighted average of the contributions of all pixels from the original
exposures that (at least partially) overlap the bin.
For each contribution, the weights are proportional to the overlap
and inversely proportional to the variance. 
The co-adding recipe also monitors the SNR and the error statistics of
the co-added spectrum, comparing the latter to the RMS of the contributions,
and raises a flag when a significant discrepancy occurs,
performing a kappa-sigma clipping (at a 3.5 $\sigma$ level)
to amend any remaining defect in the reduced spectra.

The  final spectrum covers the wavelength range $\lambda\lambda\,
305-702$ nm  
in the vacuum-heliocentric reference system. The SNR - per $R=45000$
resolution element - is $120-500$ and $ 320-500$ in the
\OVI/\Lya\ region and in the \CIV\ region (outside the \Lya\ forest), respectively. The quality
of the spectrum not only fulfills but exceeds the goals that we
proposed for the project, allowing us investigate even lower
metallicity gas than expected, as we will see in the following
sections.    


\begin{table}
\caption{Parameters obtained from the fit of the detections of \OIV\ doublets in the range $2.4902 \lsim z \lsim 3.0932$.} 
\begin{center}
\label{tab:OVIlines}
\begin{tabular}{l l c c  l l}
\hline
$z_{\rm abs}$ & $\sigma_{z}$ & $b$ & $\sigma_{b}$ & $\log N$ & $\sigma_{\rm logN}$ \\
 & & \kms & \kms & & \\
\hline
2.857951  & 0.000005 & 6.6 & 0.6 & 12.44 & 0.03 \\    
2.898767 & 0.00003  & 4.1 & 0.5 &  12.39 & 0.03 \\  
3.024823 & 0.00001 & 8.2 & 1.2 & 12.38 & 0.04 \\
\hline
\end{tabular}
\end{center}
\end{table}

\begin{table}
\caption{Parameters obtained from the fit of all detected \CIV\ lines
  in the range $2.2194 \lsim z \lsim 3.0932$.} 
\begin{center}
\label{tab:CIVlines}
\begin{tabular}{l l c c  l l}
\hline
$z_{\rm abs}$ & $\sigma_{z}$ & $b$ & $\sigma_{b}$ & $\log N$ & $\sigma_{\rm logN}$ \\
 & & \kms & \kms & & \\
\hline
  2.220038   &  0.000002    &   7.2   &   0.3  &  12.48 &   0.02  \\
  2.220242   &  0.000001    &   7.3   &   0.2  &  12.63 &   0.01  \\
  2.2205064  &  0.0000008   &   7.64  &   0.08  &  13.502 &   0.005  \\
  2.2206458  &  0.0000005   &   5.79  &   0.05  &  13.558 &   0.004  \\
  2.221070   &  0.000008    &  12.5   &   1.0  &  12.52 &   0.04  \\
  2.221312   &  0.000001    &   6.2   &   0.3  &  13.19 &   0.07  \\
  2.22142    &  0.00002     &  11.9   &   2.2  &  13.2 &   0.1  \\
  2.221608   &  0.000001    &   6.5   &   0.1  &  13.23 &   0.02  \\
  2.328888   &  0.000009    &   4.9   &   0.2  &  13.1 &   0.14  \\
  2.328957   &  0.000005    &   5.0   &   0.5  &  13.40 &   0.07  \\
  2.3291776  &  0.0000006   &  11.9   &   0.1  &  14.390 &   0.003  \\
  2.3294592  &  0.0000005   &   7.89  &   0.08  &  13.908 &   0.003  \\
  2.329641   &  0.000007    &   3.9   &   1.2  &  11.8 &   0.11  \\
  2.329929   &  0.000002    &   5.5   &   0.1  &  13.20 &   0.02  \\
  2.330057   &  0.000004    &   7.5   &   0.9  &  13.21 &   0.09  \\
  2.330294   &  0.000002    &   9.0   &   0.9  &  13.7 &   0.16  \\
  2.33049    &  0.00009     &  19.8   &  10.7  &  13.7 &   0.27  \\
  2.330669   &  0.000002    &   5.1   &   0.4  &  13.24 &   0.05  \\
  2.33071    &  0.00002     &  15.9   &   1.1  &  13.96 &   0.14  \\
  2.33077    &  0.00001     &  61.3   &   4.7  &  13.7 &   0.12  \\
  2.331130   &  0.000004    &  14.6   &   1.0  &  13.49 &   0.07  \\
  2.331239   &  0.000001    &   5.6   &   0.3  &  12.95 &   0.04  \\
  2.331482   &  0.000003    &   5.3   &   0.6  &  12.38 &   0.06  \\
  2.331623   &  0.000008    &   5.6   &   1.3  &  11.96 &   0.12  \\
  2.332278   &  0.000007    &  12.5   &   1.1  &  12.08 &   0.03  \\
  2.33257    &  0.00001     &   5.0   &   1.6  &  11.41 &   0.08  \\
  2.392362   &  0.000005    &  14.1   &   0.7  &  12.11 &   0.02  \\
  2.408887   &  0.000002    &   7.6   &   0.4  &  12.89 &   0.06  \\ 
  2.408994   &  0.000004    &  13.8   &   0.2  &  13.46 &   0.02  \\
  2.428289   &  0.000006    &  22.9   &   0.7  &  12.44 &   0.01  \\
  2.45022    &  0.00001     &   9.6   &   1.9  &  11.67 &   0.06  \\
  2.45057    &  0.00001     &   9.9   &   1.5  &  11.80 &   0.04  \\
  2.45090    &  0.00001     &   5.2   &   2.1  &  11.37 &   0.09  \\
  2.51623    &  0.00001     &  32.6   &   1.4  &  12.59 &   0.02  \\ 
  2.516660   &  0.000001    &   9.2   &   0.2  &  12.53 &   0.01  \\
  2.52909    &  0.00001     &   6.2   &   2.0  &  11.51 &   0.08  \\
  2.566401   &  0.000005    &   6.0   &   0.7  &  11.97 &   0.03  \\
  2.59645    &  0.00002     &  15.9   &   2.3  &  11.74 &   0.05  \\
  2.613624   &  0.000002    &   9.3   &   0.3  &  12.52 &   0.01  \\
  2.61389    &  0.00001     &   8.6   &   0.9  &  11.81 &   0.03  \\
  2.642776   &  0.000008    &  14.9   &   1.0  &  12.48   &   0.02 \\
  2.643171   &  0.000002    &  11.2   &   0.3  &  12.984   &   0.009  \\
  2.64364    &  0.00001     &  22.2   &   1.8  &  12.48   &   0.03  \\
  2.657911   &  0.000002    &  14.2   &   0.2  & 13.26  &  0.01 \\ 
  2.657949   &  0.000001    &   4.9   &   0.3  & 12.84  &  0.02 \\ 
  2.659252   &  0.000006    &   6.9   &   0.8  & 11.93  &  0.03 \\ 
  2.66718    &  0.00004     &  11.4   &   2.1  & 12.6  &  0.17 \\   
  2.66744    &  0.00003     &  12.1   &   3.6  & 13.1  &  0.19 \\   
  2.667632   &  0.000015    &   9.7   &   0.9  & 13.2  &  0.12 \\   
  2.667915   &  0.000002    &  12.0   &   0.2  & 13.374  &  0.008 \\   
  2.71167    &  0.000015    &  10.    &   2.   & 11.61      &  0.06 \\ 
  2.771388   &  0.000004    &   9.6   &   0.5  & 12.41      &  0.02 \\
  2.78594    &  0.00001     &  10.73  &   1.2  & 11.77     &  0.04  \\
  2.810878   &  0.000003    &  12.23  &   0.4  & 12.49      &  0.01 \\
\hline
\end{tabular}
\end{center}
\end{table}

\begin{table}
\contcaption{Parameters obtained from the fit of all detected \CIV\ lines
 in the range $2.2194 \lsim z \lsim 3.0932$.} 
\begin{center}
\label{tab:CIVlines_cont}
\begin{tabular}{l l c c l l}
\hline
$z_{\rm abs}$ & $\sigma_{z}$ & $b$ & $\sigma_{b}$ & $\log N$ &
$\sigma_{\rm logN}$ \\
 & & \kms & \kms & & \\
\hline
  2.823005   &  0.000008    &   9.8   &   1.1  & 12.02  &  0.03 \\ 
  2.82330    &  0.00002     &   6.9   &   2.5  & 11.5  &  0.13 \\ 
  2.82384    &  0.00002     &  24.7   &   2.2  & 12.81  &  0.04 \\ 
  2.82405    &  0.00001     &   5.9   &   1.8  & 11.9  &  0.23 \\ 
  2.82428    &  0.00001     &  11.0   &   1.0  & 13.07  &  0.07 \\ 
  2.8245     &  0.0001      &  11.5   &   5.7  & 12.2  &  0.46 \\ 
  2.825117   &  0.000006    &  24.8   &   1.0  & 13.00  &  0.01 \\ 
  2.825149   &  0.000002    &   6.2   &   0.5  & 12.37  &  0.04 \\ 
  2.825459   &  0.000007    &   4.4   &   1.4  & 11.7  &  0.12 \\ 
  2.825924   &  0.000006    &  10.1   &   1.1  & 12.22  &  0.08 \\ 
  2.82634    &  0.00006     &  29.8   &   6.5  & 12.45  &  0.09 \\ 
  2.8265569  &  0.0000005   &   6.5   &   0.1  & 13.188  &  0.006 \\ 
  2.826765   &  0.000005    &   5.0   &   0.9  & 11.94  &  0.08 \\ 
  2.827369   &  0.000004    &   7.4   &   0.6  & 12.29  &  0.04 \\ 
  2.827716   &  0.0000015   &   5.5   &   0.3  & 12.60  &  0.02 \\ 
  2.8280270  &  0.0000008   &  11.3   &   0.1  & 13.38  &  0.01 \\ 
  2.828064   &  0.000008    &  51.5   &   1.0  & 13.630  &  0.009 \\ 
  2.828344   &  0.000002    &   7.4   &   0.3  & 12.72  &  0.02 \\ 
  2.828868   &  0.000004    &   9.1   &   0.3  & 12.77  &  0.02 \\ 
  2.828939   &  0.000008    &   2.5   &   2.1  & 11.6  &  0.25 \\ 
  2.832878   &  0.000007    &  17.7   &   0.7  & 12.74  &  0.01 \\ 
  2.833322   &  0.000006    &  14.4   &   0.7  & 12.62  &  0.02 \\ 
  2.833965   &  0.000002    &   5.3   &   0.5  & 12.44  &  0.07 \\ 
  2.833976   &  0.000003    &  16.4   &   0.5  & 13.21  &  0.01 \\ 
  2.83451    &  0.00006     &  13.8   &   1.0  & 13.7  &  0.32 \\ 
  2.834557   &  0.000008    &   8.8   &   1.1  & 13.7  &  0.26 \\ 
  2.834810   &  0.000005    &   8.0   &   0.7  & 13.55  &  0.07 \\ 
  2.834995   &  0.000006    &   7.9   &   0.7  & 13.26  &  0.09 \\ 
  2.83511    &  0.00004     &  17.9   &   2.1  & 13.2  &  0.11 \\ 
  2.86075    &  0.00003     &   7.9   &   1.7  & 12.2  &  0.18 \\ 
  2.86091    &  0.00001     &   7.8   &   0.9  & 12.56  &  0.08 \\ 
  2.883528   &  0.000004    &  11.0   &   0.4  & 12.25  &  0.01 \\ 
  2.89872    &  0.00002     &  12.1   &   2.5  & 11.57  &  0.07 \\ 
  2.916683   &  0.000006    &   3.3   &   1.   & 11.7   &  0.1   \\
  2.916856   &  0.00002     &  22.1   &   1.   & 13.19  &  0.03  \\
  2.917109   &  0.000002    &  11.1   &   0.6  & 12.93  &  0.06  \\
  2.917555   &  0.000013    &   2.6   &   2.6  & 11.26  &  0.12  \\
  2.918023   &  0.000004    &   6.1   &   0.7  & 11.98  &  0.03  \\
  2.930791   &  0.000015    &  12.4   &   1.2  & 12.36  &  0.05  \\ 
  2.931062   &  0.000017    &  10.5   &   1.3  & 12.19  &  0.07  \\
  2.937147   &  0.000008    &  20.    &   1.   & 12.33  &  0.02  \\
  2.937751   &  0.000001    &  10.5   &   0.2  & 12.717 &  0.007 \\
  2.939641   &  0.000001    &   4.7   &   0.1  & 12.37  &  0.006 \\ 
  2.940080   &  0.000005    &  13.1   &   0.7  & 12.18  &  0.02 \\ 
  2.940455   &  0.000004    &   8.8   &   0.5  & 12.11  &  0.02 \\  
  2.95046    &  0.000015    &   9.0   &   1.7  & 11.49  &  0.06     \\
  2.98251    &  0.00002     &  12.1   &   2.4  & 11.46  &  0.06 \\
  3.024783   &  0.000004    &   4.8   &   0.7  & 11.76  &  0.04 \\ 
  3.038220   &  0.000004    &   5.9   &   0.5  & 11.83  &  0.02 \\ 
  3.038593   &  0.000001    &   9.4   &   0.1  & 12.606  &  0.005 \\ 
\hline
\end{tabular}
\end{center}
\end{table}

\begin{figure*}
\begin{center}
\includegraphics[width=8.5cm]{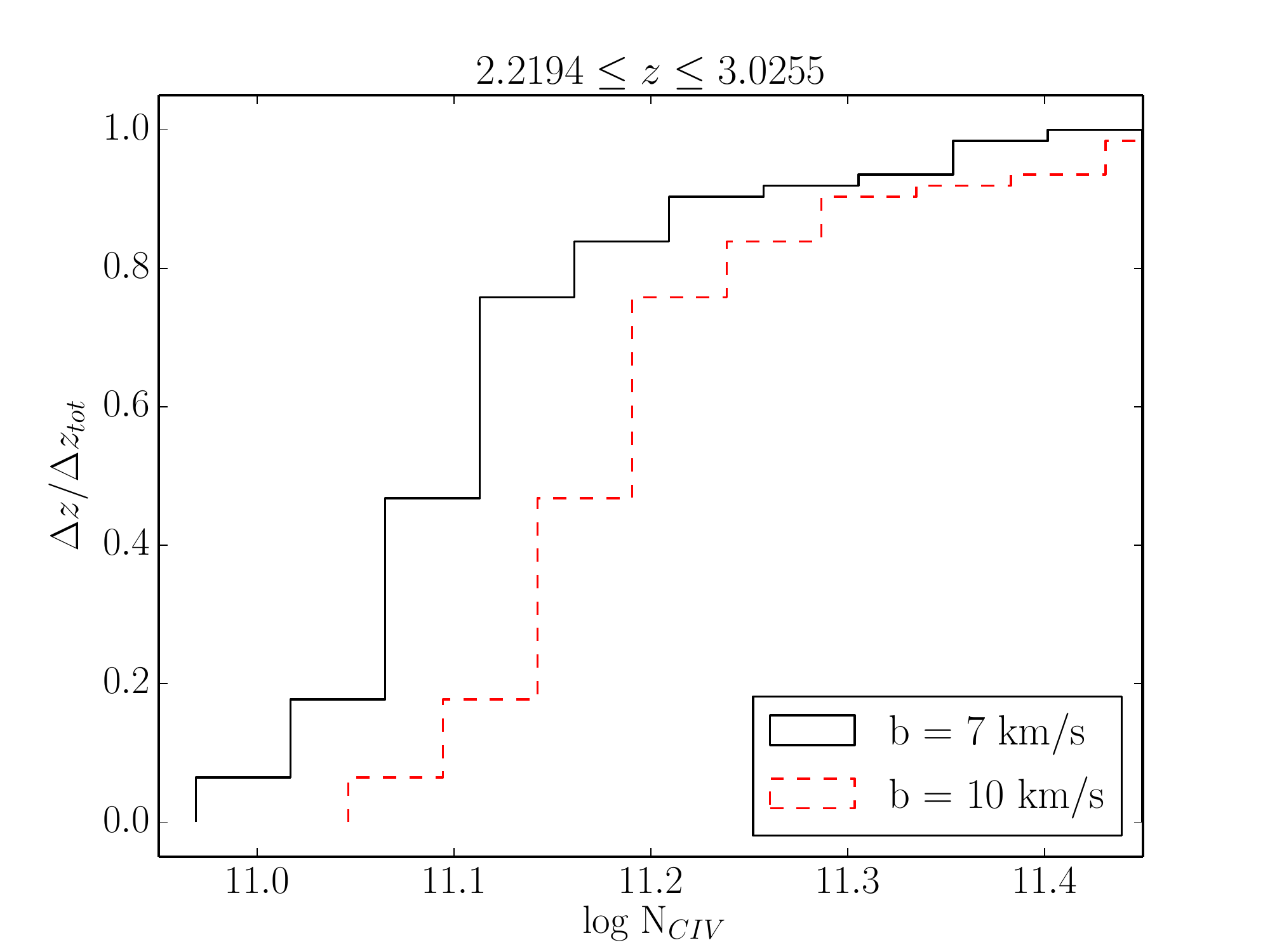}
\includegraphics[width=8.5cm]{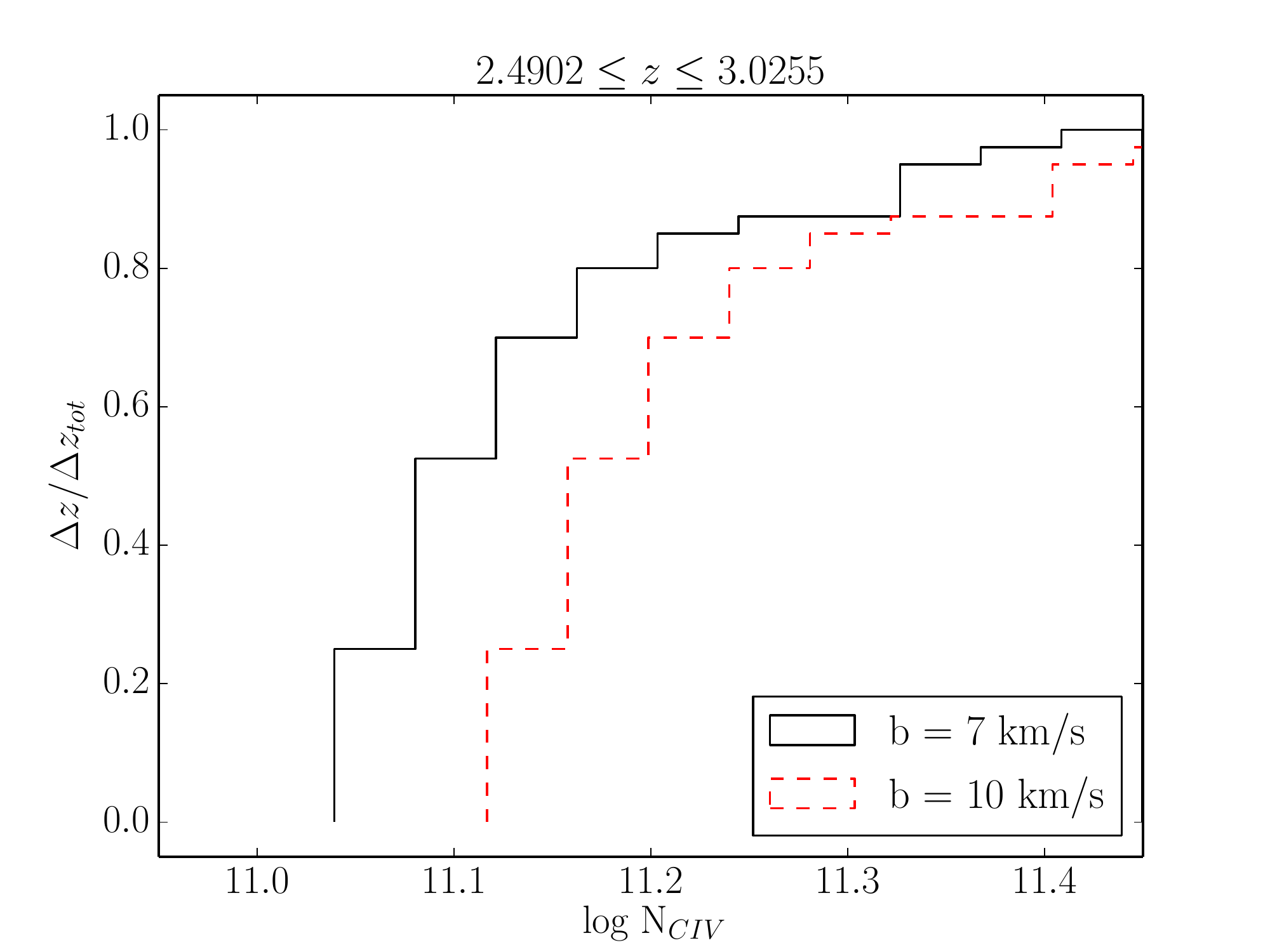}
\caption{ Cumulative fraction of two redshift ranges in the \CIV\ forest (left-hand panel: $2.2194 \lsim z \lsim3.0255$  and  right-hand panel: $2.4902\lsim z\lsim3.0255$) for which
  the deep spectrum is sensitive at $3\ \sigma$ to \CIV\ $\lambda\, 1548$ \AA\ lines
  with a column density down to the reported limits (see
  Table~\ref{tab:sensitiv}). The  sensitivity limits were computed assuming a Doppler parameter
  $b=7.0$ \kms\ (solid black line) and $b=10.0$ \kms\ (dashed red line). The definition of the two considered redshift ranges is
  reported in the text.}    
\label{fig:sensitiv}
\end{center}
\end{figure*}

\begin{table}
\caption{Fraction of the \CIV\ reported redshift range which is sensitive to
  \CIV\ $\lambda,1548$ \AA\ lines with column densities down to the
  reported limits. The values are computed for Doppler 
  parameters of 7 and 10 \kms.} 
\begin{center}
\label{tab:sensitiv}
\begin{tabular}{ c c c}
\hline
\multicolumn{3}{c}{$2.2194 \lsim z \lsim 3.0255$} \\
$\log N_{\rm CIV}$ &  \multicolumn{2}{c}{$\Delta z / \Delta z_{\rm tot}$} \\
 & $b=7.0$ & $b=10.0$ \\
\hline
10.90 &  0.000 &  0.000   \\
10.95 &  0.024 &  0.000   \\
11.00 &  0.067 &  0.008   \\
11.05 &  0.258 &  0.032   \\
11.10 &  0.657 &  0.121   \\
11.15 &  0.812 &  0.437   \\
11.20 &  0.848 &  0.725   \\
11.25 &  0.920 &  0.828   \\
11.30 &  0.928 &  0.848   \\
11.35 &  0.976 &  0.920   \\
11.40 &  1.0000 &  0.953   \\
11.45 &  1.0000 &  0.976   \\
11.50 &  1.0000 &  1.0000   \\
11.55 &  1.0000 &  1.0000   \\
\hline
\multicolumn{3}{c}{$2.4902  \lsim z \lsim 3.0255$} \\
$\log N_{\rm CIV}$ &  \multicolumn{2}{c}{$\Delta z / \Delta z_{\rm tot}$} \\
 &  $b=7.0$ & $b=10.0$ \\
\hline
10.90 &  0.000 &  0.000   \\
10.95 &  0.000 &  0.000   \\
11.00 &  0.024 &  0.000   \\
11.05 &  0.217 &  0.000   \\
11.10 &  0.608 &  0.093   \\
11.15 &  0.747 &  0.386   \\
11.20 &  0.771 &  0.639   \\
11.25 &  0.879 &  0.771   \\
11.30 &  0.891 &  0.771   \\
11.35 &  0.964 &  0.879   \\
11.40 &  1.0000 &  0.930   \\
11.45 &  1.0000 &  0.964   \\
11.50 &  1.0000 &  1.0000   \\
11.55 &  1.0000 &  1.0000   \\
\hline
\end{tabular}
\end{center}
\end{table}

\subsection{Detection and fit of metal and Hydrogen lines}

\CIV\ doublets in our deep spectrum were looked for by eye in the
region redwards of the \Lya\ emission plus 1000 \kms\ and up to 5000 \kms\ from the
\CIV\ emission to avoid contamination from the \Lya\ forest and the
proximity effect from the quasar, respectively. 
This spectral portion corresponds to the redshift range $2.2194 \lsim z \lsim
3.0255$. 

 In Fig.~\ref{fig:sensitiv} (left-hand panel) we report the cumulative fraction of
  the above redshift range which is sensitive at $3\ \sigma$
  level to \CIV\ $\lambda\, 1548$ \AA\ lines  with a column density
  down to the given values. The column density limits were
  computed using equations~(\ref{eq:ewlim}) and (\ref{eq:cdlim}) assuming a
  Doppler width of $b=7.0$ \kms\ (solid line) and $b=10.0$ \kms\ (dashed
    line) and computing the spectrum SNR in bins of 10 \AA. The adopted $b$ parameters are the mean observed value (see the end of this section) and the typical value assumed as a threshold to distinguish between Hydrogen and metal absorption lines, respectively. All the sensitivity values are reported in Table~\ref{tab:sensitiv}. 
Considering the case with $b=7.0$ \kms, the whole redshift range is
sensitive to lines with column densities down to $\log N_{\rm CIV} \simeq 11.4$. However, $\sim93$
per cent  of the considered redshift range is sensitive to lines down
to $\log N_{\rm CIV} \simeq 11.3$, the less sensitive region 
corresponds to the gap in the UVES red arm for the 564 nm setup
($2.6171\le z \le 2.6817$). Furthermore, $\sim66$ per cent of the
redshift path is sensitive to lines down to $\log N_{\rm CIV} \simeq 11.1$.

 In order to identify \CIV\ doublets we used as a first guess the
list of  detected lines in the UVES LP spectrum by \citet{dodorico10}.
Then we carried out a blind survey searching for more \CIV\ doublets
and finally, we looked for the weakest systems taking  into account
their association with \Lya\ lines. We detected 
102 \CIV\  $\lambda\,1548$ lines, of which 32 are new detections
with respect to the LP starting list. It is interesting to note that
of the 32 new detections, 15 are weak isolated lines, which are the
main target of this study.       

\begin{figure}
\begin{center}
\includegraphics[width=9cm]{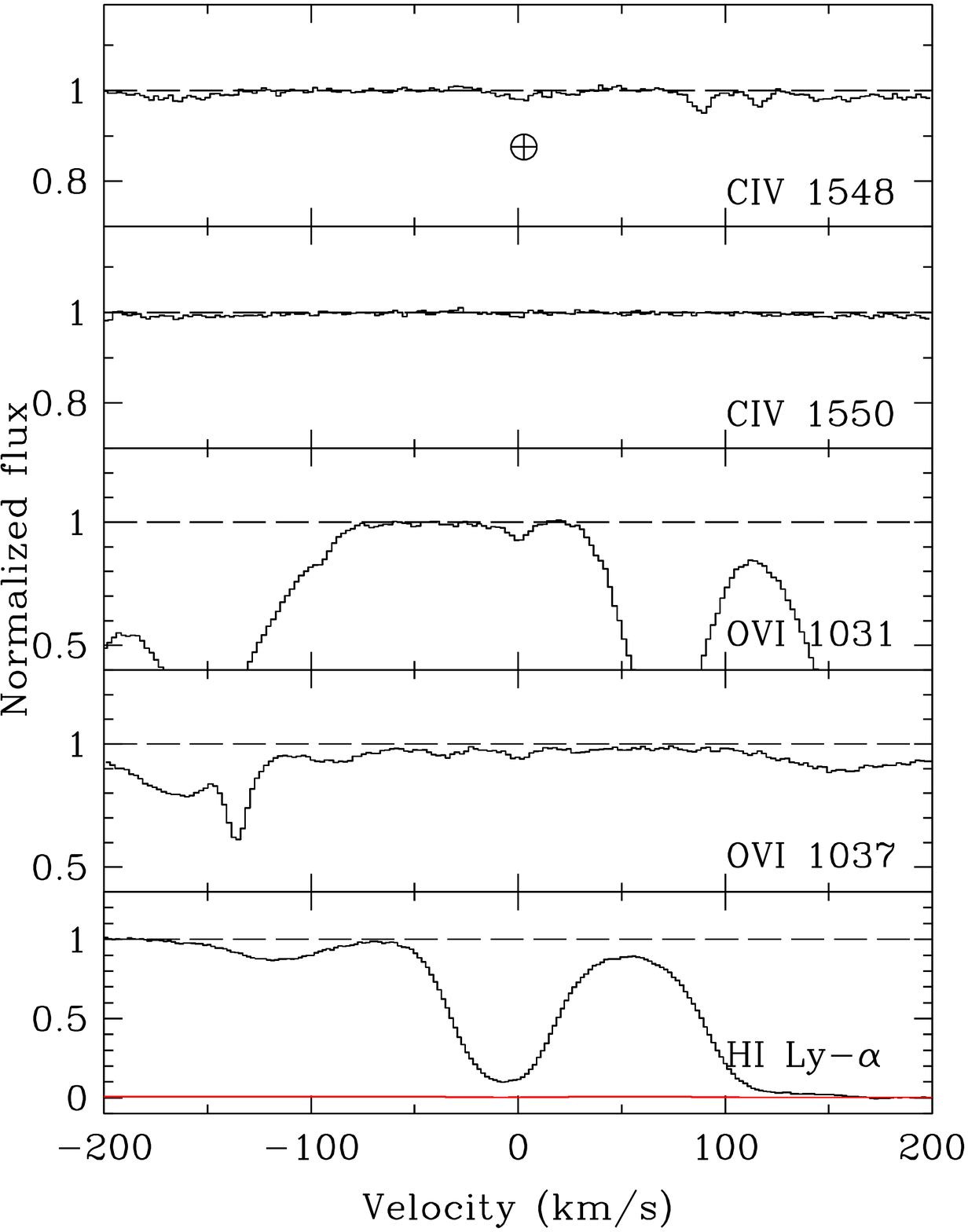}
\caption{Plot of the absorption system at $z_{\rm abs} \simeq
  2.857951$ (redshift determined by \OVI\ and corresponding to $v=0$ \kms). A weak telluric line falls at the wavelength of the
  possible  \CIV\ $\lambda\, 1548$ absorption. Note that in the panels
  showing \CIV\ $\lambda\lambda\, 1548, 1550$ \AA\ and
  \OVI\ $\lambda\lambda\, 1032, 1037$ \AA\ the y axis is zoomed to see
better the weak absorptions.}    
\label{fig:sys_z2p85}
\end{center}
\end{figure}

\begin{figure}
\begin{center}
\includegraphics[width=9cm]{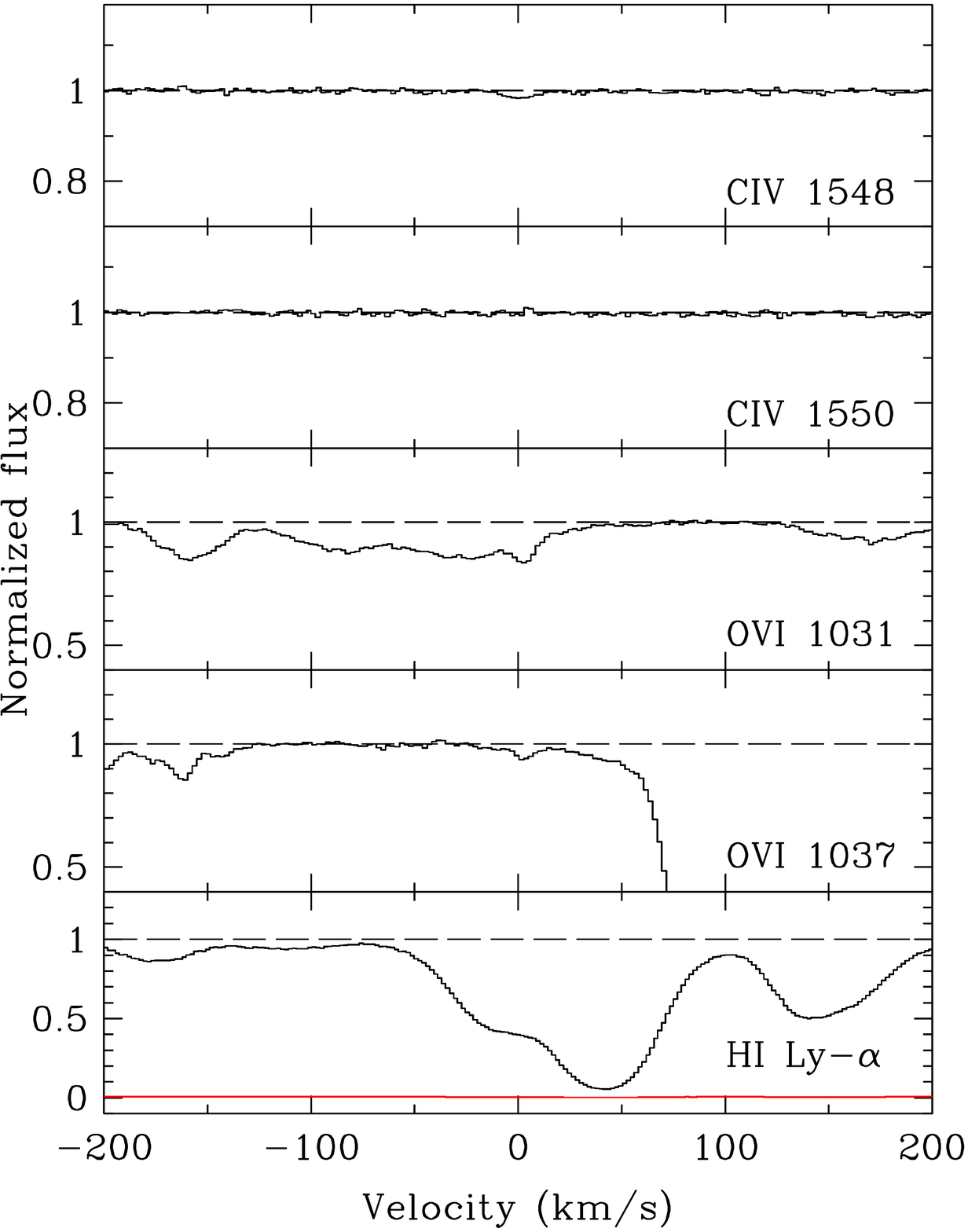}
\caption{Plot of the absorption system at $z_{\rm abs} \simeq 2.898765$ (redshift determined by \OVI).  Note that in the panels  showing \CIV\ $\lambda\lambda\, 1548, 1550$ \AA\ and \OVI\ $\lambda\lambda\, 1032, 1037$ \AA\ the y axis is zoomed to see better the weak absorptions.}   
\label{fig:sys_z2p89}
\end{center}
\end{figure}

\begin{figure}
\begin{center}
\includegraphics[width=9cm]{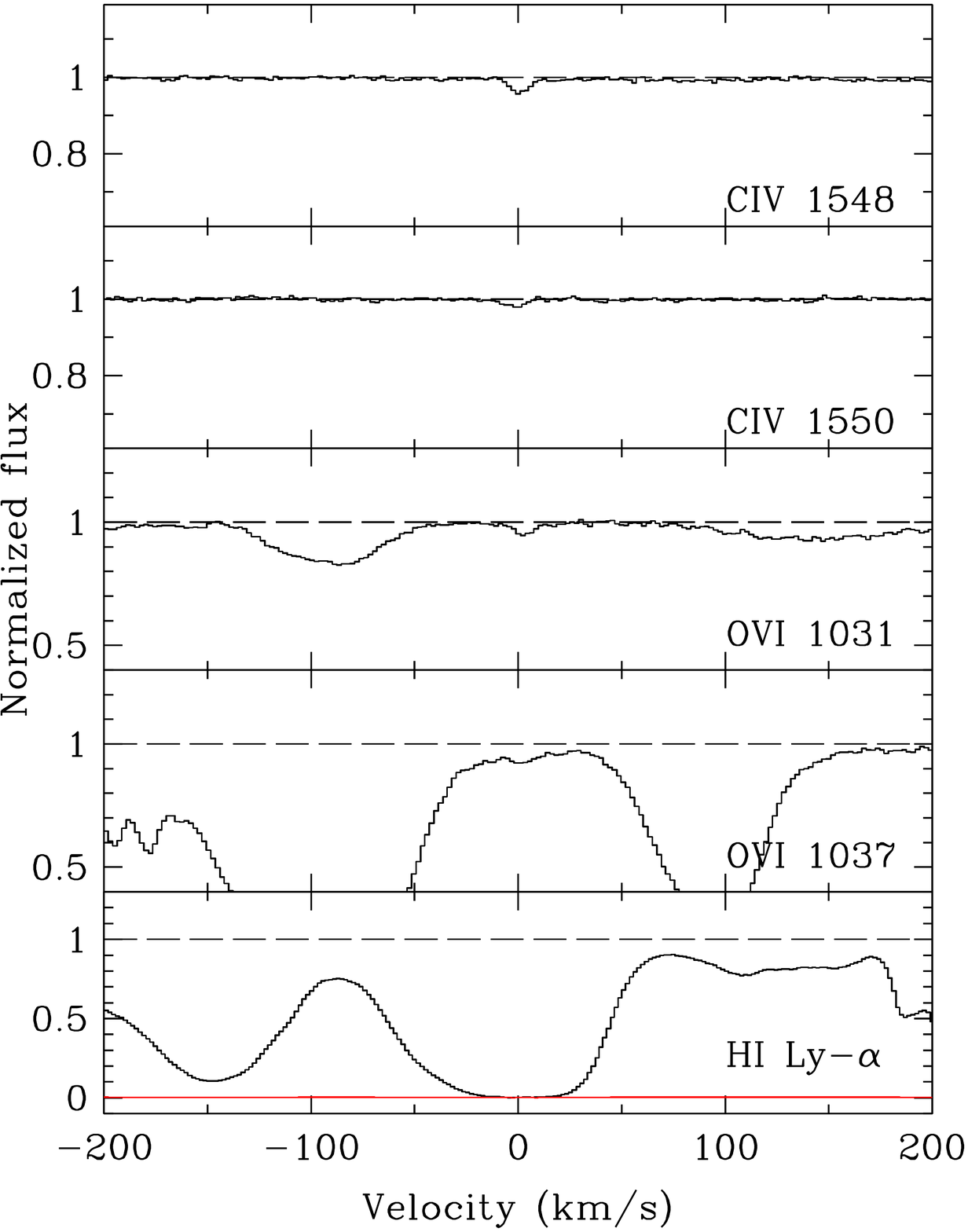}
\caption{Plot of the absorption system at $z_{\rm abs} \simeq 3.024783$ (redshift determined by \CIV).  Note that in the panels showing \CIV\ $\lambda\lambda\, 1548, 1550$ \AA\ and
  \OVI\ $\lambda\lambda\, 1032, 1037$ \AA\ the y axis is zoomed to see better the weak absorptions.}   
\label{fig:sys_z3p02}
\end{center}
\end{figure}

The \HI\ lines were identified and fitted independently of their association with metal
lines. To obtain reliable \HI\ column densities, the analysis was
focused on the redshift range for which we had at least the
\HI\ \Lya\ and \Lyb\ transitions, corresponding to $2.4902 \lsim z \lsim 3.0255$.  
 Metal contaminants in the \Lya\ forest were sought by: 
\begin{itemize}
\item [i)] including lines from ions found redwards of the \Lya\ emission line,
notably \SiII\ and \FeII;
\item [ii)] including other ions in systems found redwards of the \Lya\ emission line, e.g \CII\ if \SiII\ is present, \SiIII\ and \SiIV\ if \CIV\ is present;
\item [iii)] identifying doublets of \CIV, \SiIV, \MgII, and anything associated with them, especially for narrow lines;
\item [iv)] treating a feature as something other than \Lya\ if the corresponding
\Lyb\ feature was too weak to be consistent with the oscillator strengths.
\end{itemize}


Hereafter, the metallicity analysis will be carried out in the reduced
redshift range $2.4902 \lsim z \lsim 3.0255$. As shown in Fig.~\ref{fig:sensitiv} (right panel), $\sim60$ per cent of
this interval is sensitive to \CIV\ lines with  $\log N_{\rm CIV}$
down to 11.1, $\sim 90$ per cent to lines down to $\log N_{\rm CIV}
\simeq 11.3$ and in the whole interval we can detect at $3\ \sigma$
\CIV\ lines with column density down to $11.4$ (these values are
valid for a Doppler parameter $b=7.0$ \kms, see
Table~\ref{tab:sensitiv}). The total number of detected
\CIV\ $\lambda\,1548$ falling in the reduced redshift range is 69.   

Simulations \citep[e.g.][]{cen_chisari11,shen13} predict that \OVI\  is
a better tracer of metals than \CIV\ in the low density gas at these redshifts. 
However, due to the  location of the \OVI\ doublet ($\lambda\,\lambda 1032, 1037$ \AA) in the \Lya/\Lyb\ forests, its identification becomes extremely challenging. 
We have carried out: first a search for the \OVI\ lines associated with
the already detected \CIV\ absorptions, and  then a search to
find the weakest \OVI\ features using the association with \HI\ \Lya\ absorptions.
We considered as {\it detections} only those for which both components
of the \OVI\ doublet were identified. However, in most cases we found only one of
the two components of the doublet. We defined the latter occurrences
as {\it possible detections}\footnote{Note that in the following analysis {\it possible detections} were  counted as  upper limits.} to distinguish them from those where we could just set an upper limit to the value of $N_{\rm OVI}$ (see
below).  
We found  three \OVI\ detections (see Figs.~\ref{fig:sys_z2p85},
\ref{fig:sys_z2p89} and \ref{fig:sys_z3p02}, and Table~\ref{tab:OVIlines}) of which two were not associated with
previously identified \CIV\ lines: the system at $z_{\rm abs} \simeq 2.898765$
shows a very weak \CIV\ $\lambda\,1548$ line, while the system at
$z_{\rm abs} \simeq  2.857951$ does not show a detectable \CIV\ absorption (there
is a weak telluric line falling exactly at this wavelength). 
Besides these systems, we have also 15 \OVI\ possible detections, of which 6
do not show an associated \CIV\ line. The plots of all
\OVI\ possible detections are shown in the Appendix. 

\begin{figure}
\begin{center}
\includegraphics[width=8.5cm]{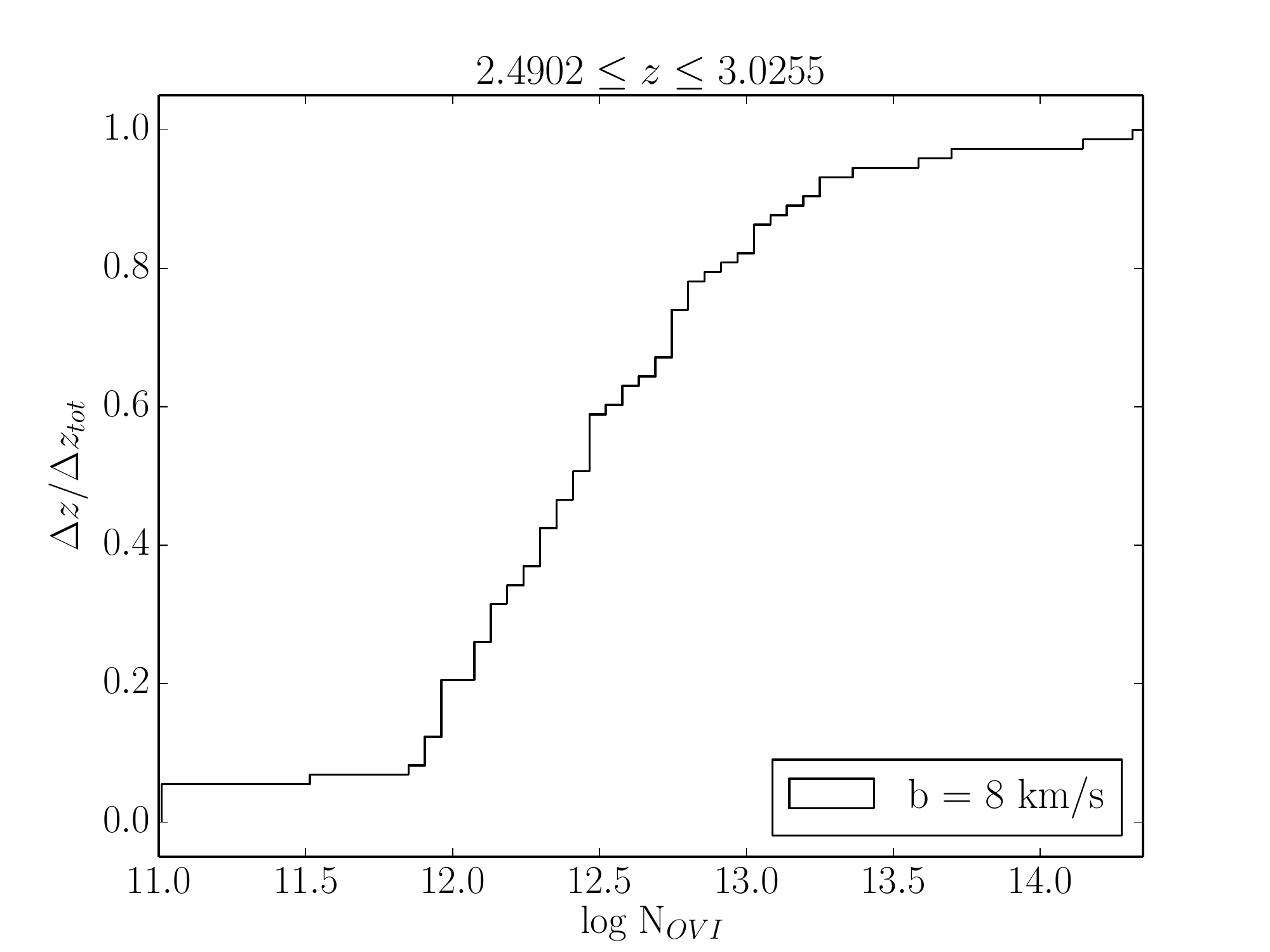}
\caption{ Cumulative fraction of the redshift range in the \OVI\ forest for which
  the deep spectrum is sensitive at $2\  \sigma$ to either line of the doublet with a column density down to the reported limits. The  sensitivity limits were computed assuming a Doppler parameter
  $b=8.0$ \kms.}    
\label{fig:sensitivOVI}
\end{center}
\end{figure}

 The determination of the \OVI\ sensitivity limits for the
  considered redshift range is extremely difficult due to the large
  number of \Lya, \Lyb\ and metal lines. We obtained an estimate of
  the $2\ \sigma$ \OVI\ limits by placing doublets with Doppler width
  $b=8$ \kms\  in the \Lya/\Lyb\ forest at a series of redshifts in
  steps of 6.7 \AA\footnote{Matching the velocity window used for the determination of the sensitivity limits in the
    \CIV\ forest.}.  Using {\small RDGEN}  \citep{rdgen}, we increased
  the column density in small steps until at least one of the lines
  was just too strong to be consistent with the data at the 2 $\sigma$ 
  level. The result is shown in Fig.~\ref{fig:sensitivOVI}.

The fit of the lines was carried out both with the Voigt profile
fitting code {\sc vpfit} \citep{vpfit} and the {\sl fitlyman}
context of the ESO {\sc midas} package \citep{font:ball}, for
comparison. 
In general, there are no significant differences between the results
of the two fits. In the following computation we adopt the results
obtained with {\sc vpfit}. 

Finally, to complete our measurement, we have determined the $3\ \sigma$ upper
limits to \CIV\ and \OVI\ column densities at the redshift of all the \HI\ lines in
our sample down to $\log N_{\rm HI} \simeq 13.5$, using formulae in
equations~(2) and (3). Doppler parameter values $b=7.0$ and
8.0 \kms\ were adopted for \CIV\ and \OVI, respectively.  Those values
are the mean Doppler parameters fitted to detected \CIV\ lines with
$\log N_{\rm CIV} < 12.0$ in our sample and to the detections and
possible detections of \OVI, respectively.  
The list of all detected \CIV\ absorption lines is reported in Table~\ref{tab:CIVlines}.

\section{Column density distribution functions}
   
The column density distribution function (CDDF), $f(N)$, is defined as
the number of lines per unit column density and per unit redshift
absorption path, $dX$ \citep{tytler87}. The CDDF is a fundamental
statistics for absorption lines, similar for many aspects to the
luminosity function for stars and galaxies.   

The redshift absorption path is used to remove the redshift dependence
in the sample and put everything on a comoving coordinate scale. In the assumed
cosmology it is defined as:

\begin{equation}
dX \equiv (1+z)^2 [ \Omega_{\rm m}(1+z)^3 + \Omega_{\Lambda}]^{-1/2} dz.
\end{equation}  

 With the adopted definition, $f(N)$ would be the same at any redshifts
for a non-evolving population of absorbers.  

\begin{figure}
\begin{center}
\includegraphics[width=9.5cm]{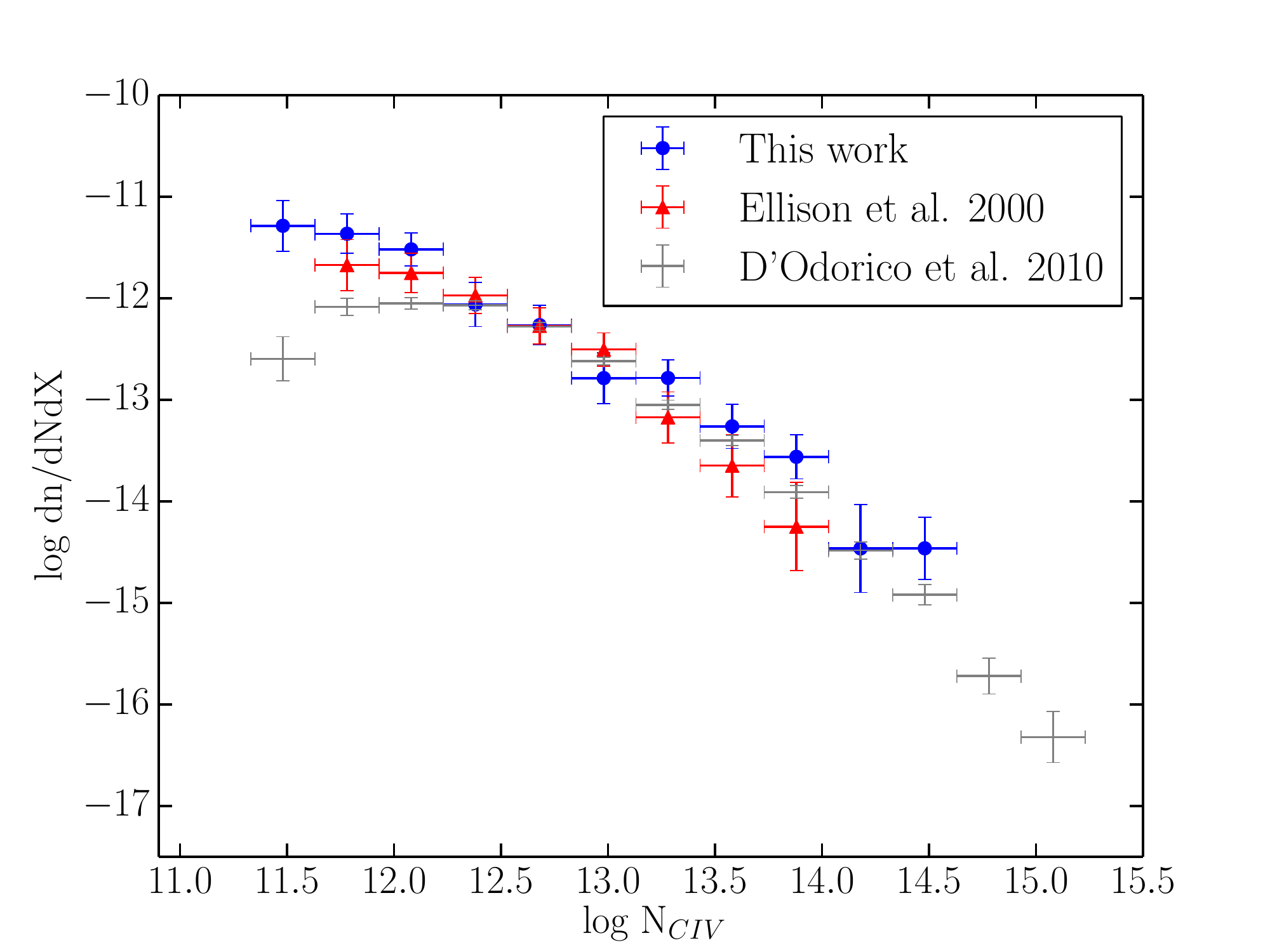}
\caption{CDDF of \CIV\ lines in the spectrum of HE0940-1050 (blue
  points). The bin-size is $10^{0.3} N_{\rm CIV}$ cm$^{−2}$ and the
  error bars are $\pm 1\ \sigma$, based on the number of points in
  each bin. Superposed are the CDDFs computed for the list of
  \CIV\ lines detected by ESSP00 (red triangles) and for the
  \CIV\ lines in \citet[][grey crosses]{dodorico10}. In the three
  samples, lines closer than 50 \kms\ have been merged (see text).}   
\label{fig:cddf_civ}
\end{center}
\end{figure}

\begin{figure}
\begin{center}
\includegraphics[width=9.5cm]{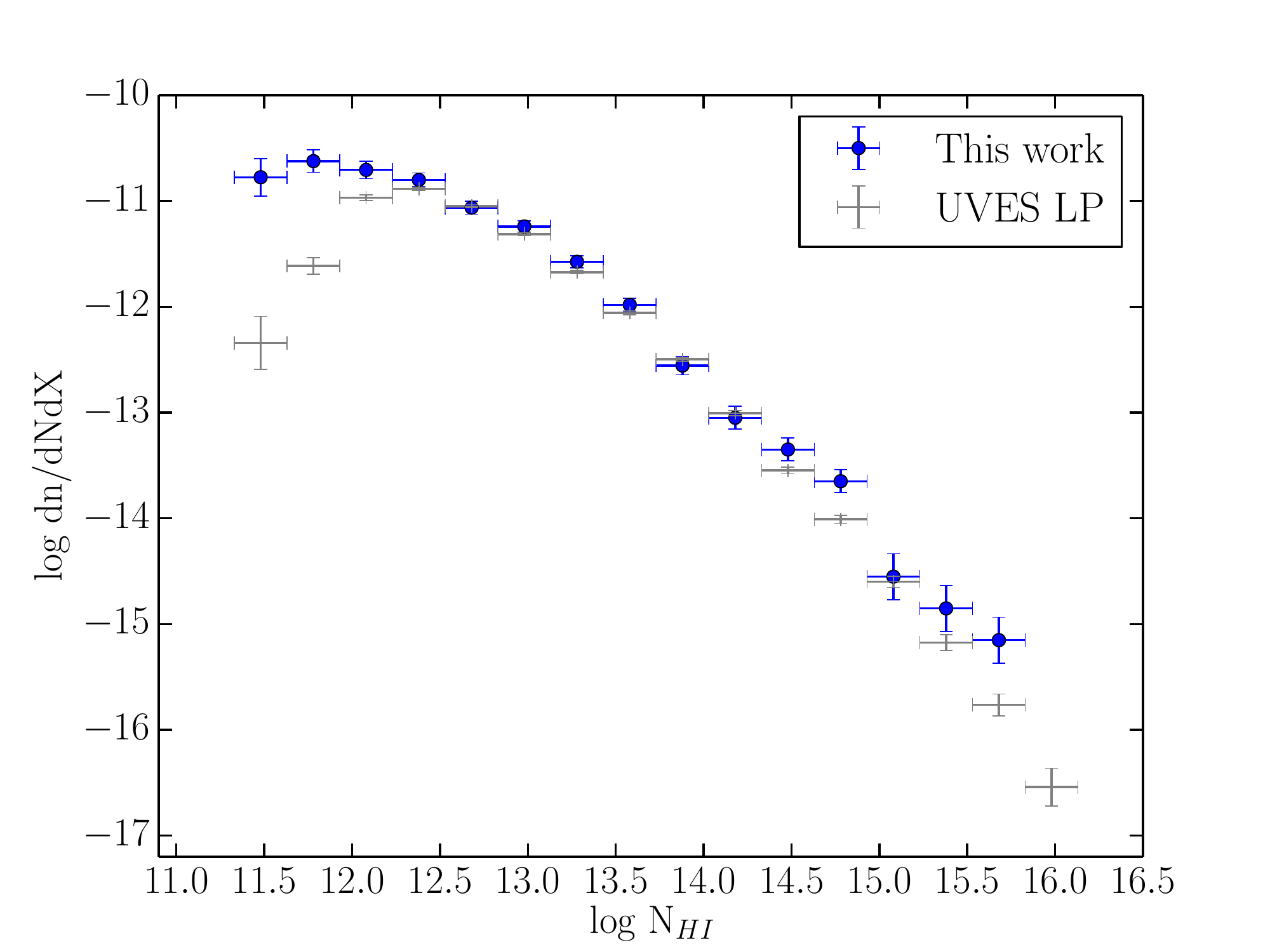}
\caption{CDDF of \HI\ lines in the spectrum of HE0940-1050 (blue
  points). The bin-size is $10^{0.3} N_{\rm HI}$ cm$^{−2}$ and the
  error bars are $\pm 1\ \sigma$, based on the 
  number of points in each bin. Superposed is the CDDFs computed for
  the \HI\ lines of the quasars in the UVES LP \citep[][grey
    crosses]{dodorico08}.}    
\label{fig:cddf_lya}
\end{center}
\end{figure}

In Fig.~\ref{fig:cddf_civ} the CDDF for the  \CIV\ lines  detected in
the deep spectrum is compared with the CDDF computed for the
\CIV\ lines detected by ESSP00\footnote{Note that data are not
  corrected for completeness. The difference in absolute values with
  respect to the original result is due to the merging of the lines
  and to the different cosmological models adopted to compute $dX$.}
and with the CDDF computed for the sample of more than 1000 lines by
\citet{dodorico10}. In the three samples, lines closer than 50
\kms\ have been merged into a single line with column density equal to
the sum of the column densities and redshift equal to the column
density weighted mean of the original redshifts. This operation was
carried out in order to avoid spurious effects in the low column
density regime due to the weak \CIV\ components introduced by the
process of fitting complex systems (note that   the three
samples were fitted with different softwares).  

It is evident from Fig.~\ref{fig:cddf_civ} that, although the three samples cover
different redshift intervals, there are no significant differences for
$\log N_{\rm CIV} \ge 12.3$. In the low column density regime, our
result confirms the result obtained by ESSP00 with the completeness
correction and extends the increasing trend down to  $\log N_{\rm CIV}
\simeq 11.3$. On the other hand, the comparison with the lower SNR
sample shows the significant improvement obtained with the deep
spectrum at the lowest column densities.  
  
The CDDF of the \HI\ lines is shown in Fig.~\ref{fig:cddf_lya}. A
comparison with the \HI\ lines fitted in the quasars of the UVES LP
shows significant differences for $\log N_{\rm HI} < 12$. In both
samples we have selected lines to have $10 \le b \le 200$ \kms\ and
errors on the column density $\sigma(N_{\rm HI}) < 0.5 N_{\rm HI}$ and on redshift
$\sigma(z) < 8 \times 10^{-5}$  (corresponding to about 3 pixels).  

\begin{figure}
\begin{center}
\includegraphics[width=9.5cm]{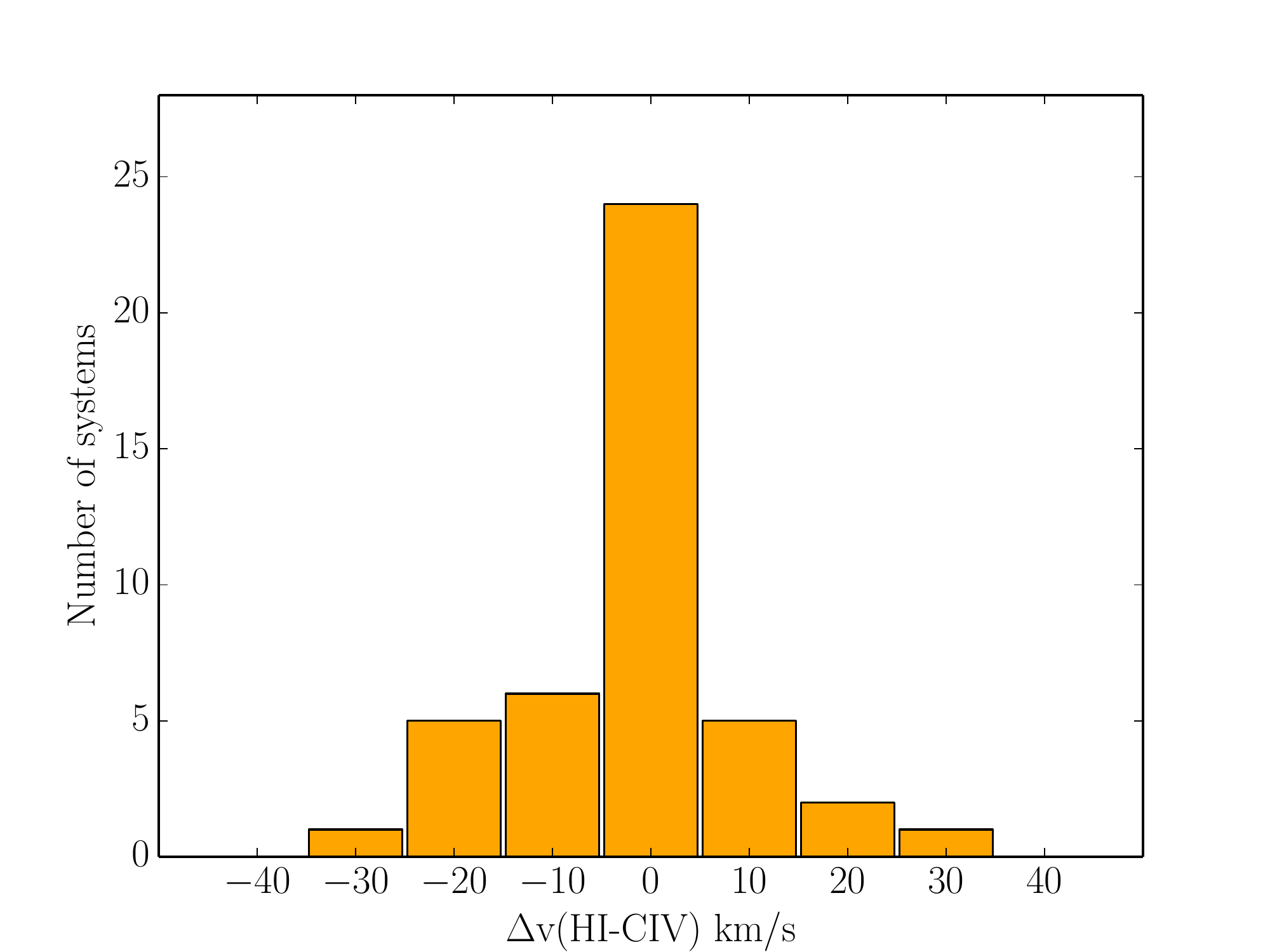}
\caption{Distribution of the velocity shifts between associated \HI\ and \CIV\ lines.}   
\label{fig:CIV_vshift}
\end{center}
\end{figure}


\begin{figure}
\begin{center}
\includegraphics[width=9.5cm]{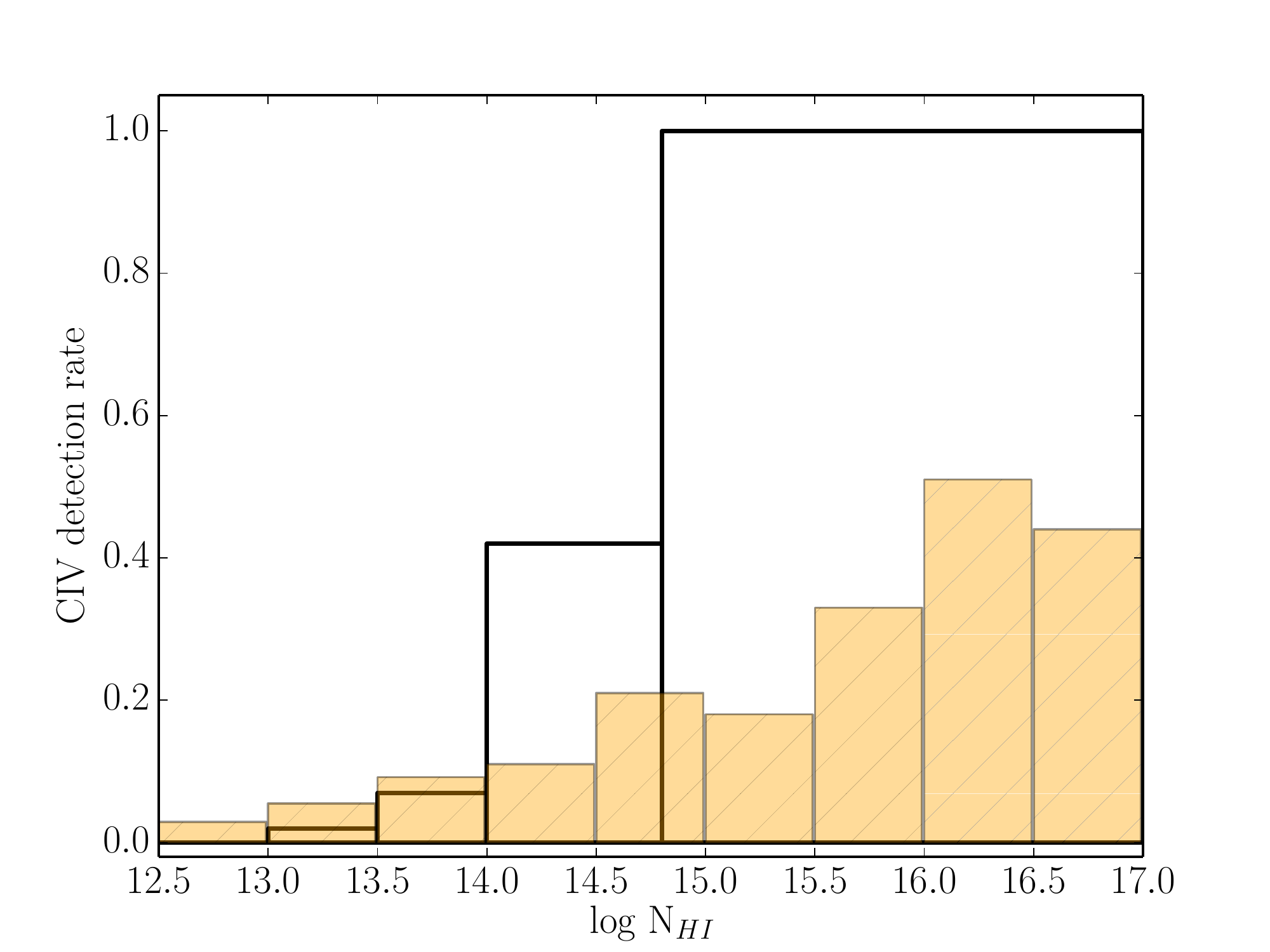}
\caption{Detection rate of \CIV\ absorbers (down to our detection
  limit of $\log N_{\rm CIV} = 11.4$) as a function of the column
  density of the associated \HI\ line (white, open histogram). The overplotted hatched
  histogram is the fraction of \HI\ systems that arise within $|\Delta\,v|
< 300$ \kms\ and $D_{\rm tran} < 300$ physical kpc from LB galaxies \citep[adapted
from][]{rudie12}.}   
\label{fig:CIV_detrate}
\end{center}
\end{figure}

\section{\CIV\ detection rate and the connection with galaxies}
The main goal of this work is to shed light on the abundance and
covering factor of metals in the low density gas. To this aim we have
computed the detection rate of \CIV\ lines (down to our observational
limit) as a function of the column density of the associated
\HI\ absorption. 

\begin{table}
\caption{\CIV\ detection rate  down to our (conservative) detection
  limit of $\log N_{\rm CIV} = 11.40$ in the spectrum of HE0940.}
\begin{center}
\label{tab:rate}
\begin{tabular}{c r r c}
\hline
$\log N_{\rm HI}$ & No. of lines & No. of det. & Det. rate \\
\hline  
$\ge 15.0$ & 16 & 16 & 1.00 \\
$[14.5:15.0]$ & 21 & 12 & 0.57 \\
$[14.0:14.5]$ & 28 & 12 & 0.43  \\
$[13.5:14.0]$ & 56 &  4 & 0.07 \\
$[13.0:13.5]$ & 98 &  2 & 0.02 \\
\hline
$\ge 14.8$ & 22 & 22 & 1.00 \\
$[14.0:14.8]$ & 42 & 18 & 0.43  \\
$[13.5:14.0]$ & 56 &  4 & 0.07 \\
$[13.0:13.5]$ & 98 &  2 & 0.02 \\
\hline
\end{tabular}
\end{center}
\end{table}

The association \HI\,-\,\CIV\ absorbers has been carried out in the
following way. First, we have determined all the pairs \HI\,-\,\CIV\ absorber at velocity separations closer than $dv = 50$
\kms. This is approximately the smallest velocity separation for which all
\CIV\ lines in our sample are associated with at least one
\HI\ line. Then, if the same \CIV\ line was associated with multiple
\HI\ lines we kept only the closest association. Finally, if more than
one \CIV\ line was associated with the same \HI\ line, we merged the
\CIV\ lines into a single system with column density equal to the sum
of the column densities and redshift equal to the column
density weighted mean of the original redshifts. The distribution of the velocity shifts between the associated \HI\ and \CIV\ lines is shown in Fig.~\ref{fig:CIV_vshift}.

Having obtained the final list of one to one associations between
\HI\ and \CIV\ lines, the \CIV\ detection rate was computed by dividing
the number of \HI\ lines with an associated \CIV\ absorption in a
given \HI\ column density bin by the total number of \HI\ lines in that bin. 
 All the computed detection rates are dependent on the completeness limit of our sample which is $\log N_{\rm CIV} = 11.4$ for the whole considered redshift range (see Section 2.2 for more details). 

Our findings (summarized in Table~\ref{tab:rate}) are consistent with
the results of the first works on metals
in the IGM \citep[e.g.][]{cowie95}: above a column density of $\log
N_{\rm HI}\simeq 15$ all \HI\ lines have associated \CIV\ lines, while
in the column density bin $14.5 \leq \log N_{\rm HI} < 15$, 57
per cent of the observed \HI\ absorptions have associated
metals. Actually, thanks to our exceptional sample, we find a
\CIV\ detection rate of 100 per cent for  $\log N_{\rm HI}\ge 14.8$,
while we can state that 43 per cent of \HI\ lines with  $14.0 \leq 
\log N_{\rm HI} < 14.8$ have an associated \CIV\ absorption. 
Below $\log N_{\rm HI} \simeq 14.0$, we observe a drop with detection
rates lower than 10 per cent. 
Of course, we have to take into account that our measurements are based
on a single spectrum and that, in particular for $\log N_{\rm HI} \lsim
14.5-15$ there could be variations from one line of sight to the other.

In Fig.~\ref{fig:CIV_detrate}, we compare our result for the
\CIV\ detection rate with the fraction of \HI\ lines which arises
within $|\Delta\,v|< 300$ \kms\ (difference in redshift) and $D_{tran}
< 300$ physical kpc (transverse separation) of star forming galaxies at $2
\lsim z \lsim 2.8$ as obtained by \citet[][]{rudie12}.
The comparison shows that the metal detection rate for lines  with $\log
N_{\rm HI} \ge 14.0$ is more than a factor of 2 larger than the fraction
of \HI\ lines tracing the circum-galactic medium (CGM) of the galaxies
in the sample used by Rudie and collaborators. Those galaxies inhabit
dark matter haloes of average mass $\sim 10^{12}$ M$_{\sun}$ and the fainter objects
correspond to $0.25L^*_{\rm UV}$ (at $\langle z \rangle = 2.30$).    

Our result points out that metals are diffused also outside
the CGM of the considered galaxies.   They could reside in filaments
enriched by {\it in situ} star formation or in the outskirts of smaller, 
fainter galaxies which are not present in the sample of
\citet{rudie12}.  Metals could also lie at larger distances than
  the 300 physical kpc assumed by Rudie and collaborators, possibly
  due to enrichment by earlier generations of galaxies. Hints of the
  fact that metals are widespread beyond the regions around galaxies
  were found also in previous observational works
  \citep[e.g.][]{pieri06,martin10}.

\section{The metallicity of the IGM}
Having computed the \CIV\ detection rate as a function of the
\HI\ column density, we can ask ourselves if the drop that we observe
below  $\log N_{\rm HI} \simeq 14.0$  in Fig.~\ref{fig:CIV_detrate} is due to a real decrease of the
metal enrichment of the IGM at those overdensities or if it could be
ascribed to  our limited sensitivity. In general, we would like to know
which is the characteristic metallicity of the CGM/IGM gas at the
investigated redshifts.
 
\begin{table}
\caption{Parameters of the \cloudy\ models. We adopted grid in
  overdensity and metallicity. Metallicity is not reported
  in the table, we considered metallicites with respect to solar $\log
  Z/Z_{\sun}=[-4.0, -3.5, -3.0, -2.5, -2.0, -1.5, -1.0]$. In each cell of the table, we report
  the total number density, the \HI\ column density (both in log) and
  the temperature for the corresponding overdensity at $z=2.8$. See text for more
  details.} 
\begin{center}
\label{tab:cloudy}
\begin{tabular}{l c c c c c c}
\hline
&  \multicolumn{6}{c}{$(\delta +1)$} \\
& 1.0  & 3.0 & 5.0 & 12.0 & 15.0 & 50.0  \\
\hline
$\log n_{\rm H}$ & $-5.02$ & $-4.55$ & $-4.33$ & $-3.846$ & $-3.95$ &
$-3.33$  \\
$\log N_{\rm HI}$ & 13.33 & 13.98 & 14.29 & 14.81 & 14.94 & 15.66  \\
 T/ $10^4$ K & 1.0 & 1.7 & 2.2 & 3.5 & 3.9 & 7.1 \\  
\hline
\end{tabular}
\end{center}
\end{table}

\begin{figure*}
\begin{center}
\includegraphics[width=18cm]{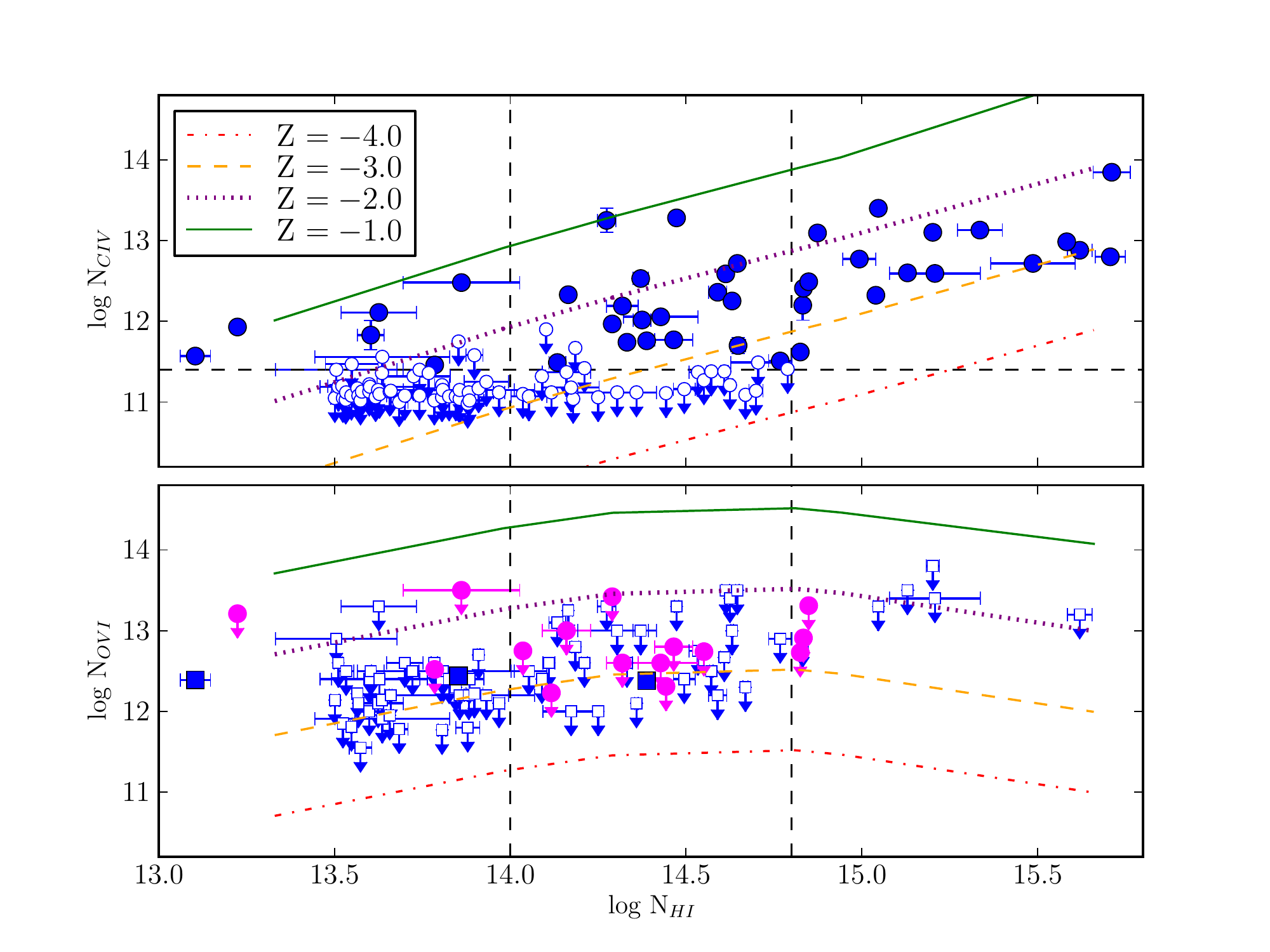}
\caption{Upper panel: measurements (blue filled dots) and upper limits
  (open dots) of  $\log N_{\rm CIV}$ as a function of the associated
  \HI\ column density for our sample of absorbers. The horizontal
  dashed black line marks the (conservative) sensitivity limit $\log
  N_{\rm CIV} = 11.4$. Lower panel:  measurements (blue filled squares), possible detections (pink filled dots) and upper
  limits (open squares) of  $\log N_{\rm OVI}$  as a function of the
  corresponding \HI\ column density. Upper limits have not been
  measured for $\log N_{\rm HI} < 13.5$. Overplotted are the
  corresponding column densities obtained from our photoionization
  models at $z=2.8$; different lines correspond to different
  metallicities as marked in the figure. The two dotted vertical lines
  have been drawn at column densities $\log N_{\rm HI}=14$ and 14.8
  corresponding at $z=2.8$ to $(\delta+1)=3.1$ and 11.8, respectively. See
  the text for more details.}    
\label{fig:CIV_OVI}
\end{center}
\end{figure*}

To this aim, we have generated a set of simple photonionization models
using version 13.0 of \cloudy, last described
by \citet{cloudy}. In the models, we have assumed a plane parallel
slab of gas at a constant temperature $T$, ionized by the inbuilt Haardt \& Madau UV background
\citep{hm01} at redshift $z=2.8$. The computation stops when a given column density of 
\HI\ is reached and the column densities of the studied ions are given
in output (solar relative abundances are assumed). We built a grid of
models varying overdensity and metallicity.  The corresponding temperature $T$ is
defined by the temperature-density relation valid for the IGM
\citep{hui_gnedin}: 
\begin{equation}
\label{eqn:TDR}
T = T_0 (1 + \delta)^{\gamma -1}
\end{equation}

\noindent
where we assume $T_0 = 10^4$ K and $\gamma = 1.5$ \citep[following][]{becker11}.
The total density is computed based on the mean density at the considered redshift and the
considered overdensity. The \HI\ column density used to stop the
computation is determined using the formula by \citet{schaye01}: 

 \begin{eqnarray}
 \label{eq:schaye}
 N({\rm HI})  & \simeq&  2.7 \times 10^{13} {\rm cm}^{-2}\, (1+\delta)^{1.5-0.26(\gamma-1)}\, T_{0,4}^{-0.26} \\  \
 \nonumber
 && \Gamma_{-12}^{-1}\,   \left( \frac{1+z}{4}\right )^{9/2}       
\end{eqnarray}

\noindent
where we have adopted $T_{0,4} \equiv T_0 / 10^4 {\rm K} = 1$,
$\Gamma_{-12} \equiv \Gamma / 10^{-12} =1$ and $\gamma = 1.5$, as above. 
All the model parameters are reported in Table~\ref{cloudy}.

In Fig.~\ref{fig:CIV_OVI}, we plot our observational results for
\CIV\ and \OVI\ versus \HI\ column densities for each absorber,
together with the predictions of the photoionization models described
above. In the plots, we empirically identify  three possible regimes. At 
$\log N_{\rm HI} \ge 14.8$ (corresponding to $(\delta + 1) \simeq 11.8$ at
$z=2.8$), we are probing regions permeated by metals with $-3 \lsim \log
Z/Z_{\sun} \lsim -1$, likely tracing the CGM close to galaxies. The
range $14.0 \le \log N_{\rm HI} < 14.8$ is where 
observationally we are most sensitive to the very low metallicity gas:
what we observe is that  $\sim 60$ per cent of absorbers are enriched
to metallicities $-3 \lsim \log Z/Z_{\sun} \lsim -1$ (considering
upper limits as measurements); while  $\sim 40$ per cent of absorbers
are characterized by metallicities $\log Z/Z_{\sun} \lsim -3$, mostly
traced by \CIV\ upper limits. Assuming that all \CIV\ upper limits would
trace metallicities $\lsim -3$, a lower limit of 38 per cent of absorbers would
have $-3 \lsim \log Z/Z_{\sun} \lsim -1$. The possible \OVI\ detections (filled pink dots) and the
single detection (filled blue square) observed in this \HI\ column
density regime trace  a slightly lower metallicity interval ($-3
  \lsim \log Z/Z_{\sun} \lsim -2$). For $\log N_{\rm HI} < 14.0$ our
measurements of \CIV\ lines are no longer sensitive to metallicities 
$\log Z/Z_{\sun} \lsim-3$ (and going down in $N_{\rm HI}$, the limit rises
above $\log Z/Z_{\sun} \sim-2$). As a consequence, even if the gas traced by those
\HI\ absorbers was enriched at the same level as the gas traced by
larger \HI\ column densities, our observations could not detect the
associated \CIV\ lines. On the other hand, more stringent upper limits
on the metallicity in this overdensity regime can be put using
\OVI\ (lower panel of Fig.~\ref{fig:CIV_OVI}) for which we find that
$\sim88$ per cent of the absorbers have metallicities $\log Z/Z_{\sun} < -2$ 
and, in particular, $\sim24$ per cent of absorbers have $ \log
Z/Z_{\sun} < -3$.    

\begin{figure}
\begin{center}
\includegraphics[width=9cm]{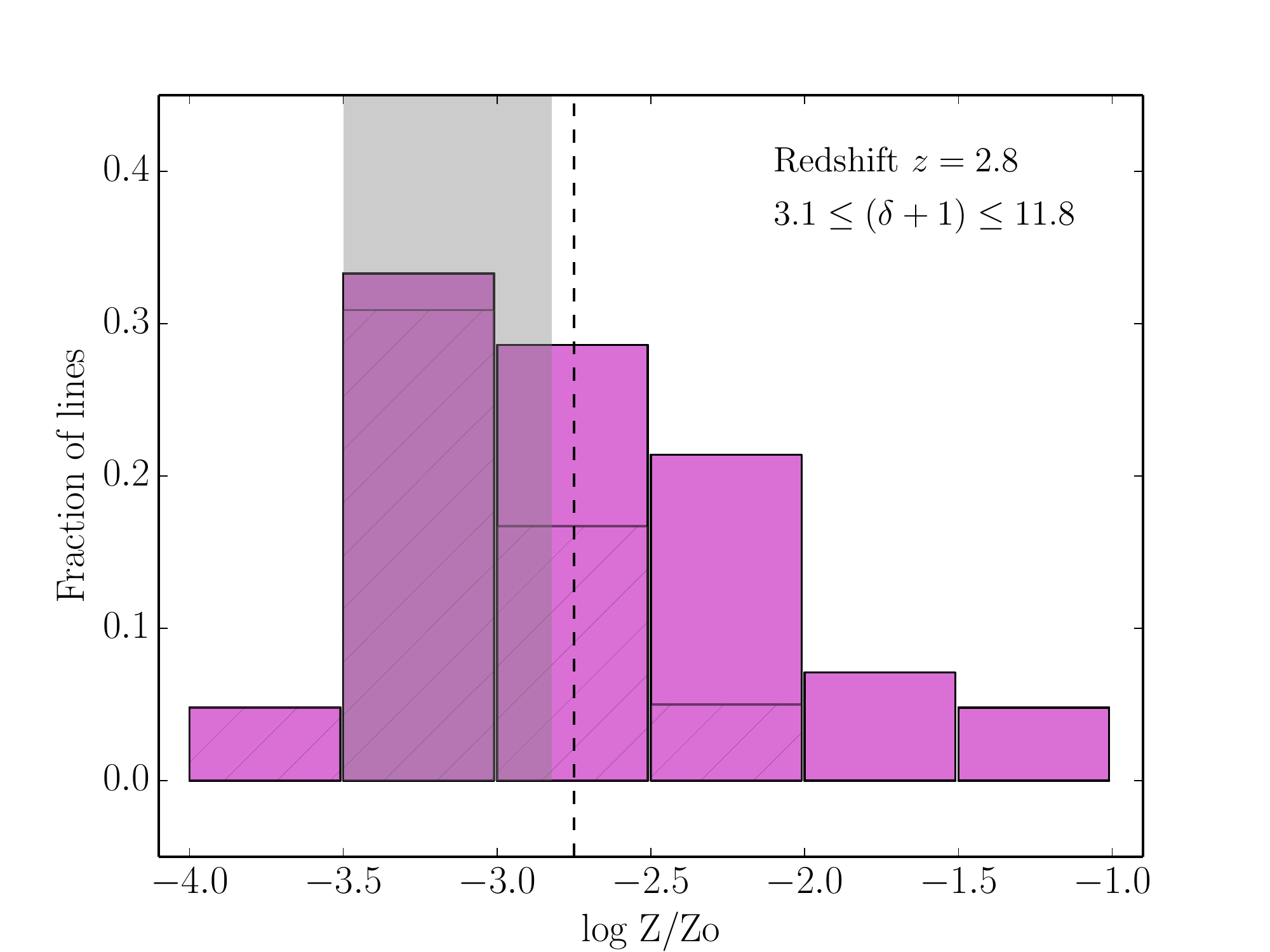}
\caption{Fraction of observed \CIV-\HI\ absorbers with $14 \le N_{\rm
    HI} \le 14.8$ falling in different metallicity bins, when compared
  with \cloudy\ predictions at $z=2.8$. The hatched 
  histograms correspond to the contribution of upper limits in each
  metallicity bin. The grey band is the prediction for the median
  [C/H] abundance computed with eq.uation~(1) by S03 for the considered
  overdensity range. See text for
  more details.}   
\label{fig:metal_hist}
\end{center}
\end{figure}

With the caveat that we are dealing with a small sample, we carried
out a more detailed analysis of the metallicity of the gas probed by
the \HI\ absorption lines with $14.0 \le \log N_{\rm HI} < 14.8$,
range in which our observations are more sensitive to the very low
metallicities. This column density interval, when translated into
overdensities with equation~(\ref{eq:schaye}), corresponds to $3.1 \le (\delta
+ 1) \le 11.8$ at $z=2.8$. 
In the considered column density interval, 42 \HI\ lines were detected, of which 18
have an associated \CIV\ absorber, while for the remaining 24 lines we have determined an
upper limit to $\log N_{\rm CIV}$. 

The sample of \HI-\CIV\ pairs has
been compared with the results of our \cloudy\ models at redshift
$z=2.8$, computed for the finer metallicity grid: $\log
Z/Z_{\sun}= [-4.0, -3.5, -3.0, -2.5, -2.0, -1.5, -1.0]$. The results
of this comparison are shown in form of histogram of the fraction of
lines falling in each metallicity bin in 
Fig. ~\ref{fig:metal_hist}. The figure shows that the
majority of the lines falls in the bin at $\log
Z/Z_{\sun}=[-3.5,-3.0]$, but the majority of the \CIV\ detections
falls in the range of metallicities $\log Z/Z_{\sun}=[-2.5,-2.0]$. The
median of the distribution of lines falls in the bin $\log
Z/Z_{\sun}=[-3.0,-2.5]$. 

\subsection{Comparison with POD results}
 
ESSP00 estimated the relation between $\tau_{\rm HI}$ and $\tau_{\rm
  CIV}$ with the POD method using the extremely high SNR spectrum of
B1422+231. Using a simple simulation, they showed that  the POD result
could be reproduced only assuming that more metals than those directly
detected were present. In particular, they assumed that all lines with
$\log N_{\rm HI} \ge 14.5$ had  an associated \CIV\ line with at
least $\log N_{\rm CIV} = 12.0$ and that all weak \Lya\ lines ($\log
N_{\rm H I} <14.5$) were enriched with a constant
$\log~$\CIV/\HI$~=-2.6$. 

We have verified the validity of those conditions for our sample,
starting from the hypothesis that the POD method is a cumulative
technique. Thus, we can compute the total column density in
\CIV\ corresponding to a certain \HI\ column density range and divide
it by the number of \HI\ lines to obtain the average \CIV\ column
density to be compared with ESSP00. First, all the \HI\ lines in our
sample with $\log N_{\rm HI} \ge 14.8$ have an associated \CIV\ line
with $\log N_{\rm CIV} \ge 12.0$.  In the range of \HI\ column densities $14.5 \le \log
N_{\rm HI} < 14.8$, we have computed the average \CIV\ column density
to be $11.97 \le \log N_{\rm CIV} \le 12.02$. The lower limit is
obtained assuming that the \HI\ lines without a \CIV\ detection have
zero \CIV\ and the upper limit considering the \CIV\ upper limits
as measurements. 
On the other hand, for the  \HI\ lines with $13.5 \le \log N_{\rm
  HI} < 14.5$ and adopting the same assumptions described above the
average ratio varies between $-2.16 \le \log ~$\CIV/\HI$~ \le -2.08$. This
implies that with our data we have directly detected the metals that
ESSP00 found statistically applying the POD method.   

In Fig.~\ref{fig:metal_hist}, our result obtained through the direct
detection (or non-detection) of 
\CIV\ lines has been compared with the median [C/H] abundance derived
from the POD analysis by S03  and computed with equation~(\ref{eq:pod}) at
the redshift and overdensities reported in the figure. The two 
median abundances are in good agreement, if we take into account all the
uncertainties derived by the rough modelling adopted in our work and 
also introduced by the conversion of the observed quantities into the [C/H]  abundance
in S03.   

The POD analysis for the spectrum of HE0940-1050 will be presented in
a forthcoming paper (D'Odorico et al., in preparation).  

\subsection{Volume filling factors}

\begin{figure}
\begin{center}
\includegraphics[width=9.5cm]{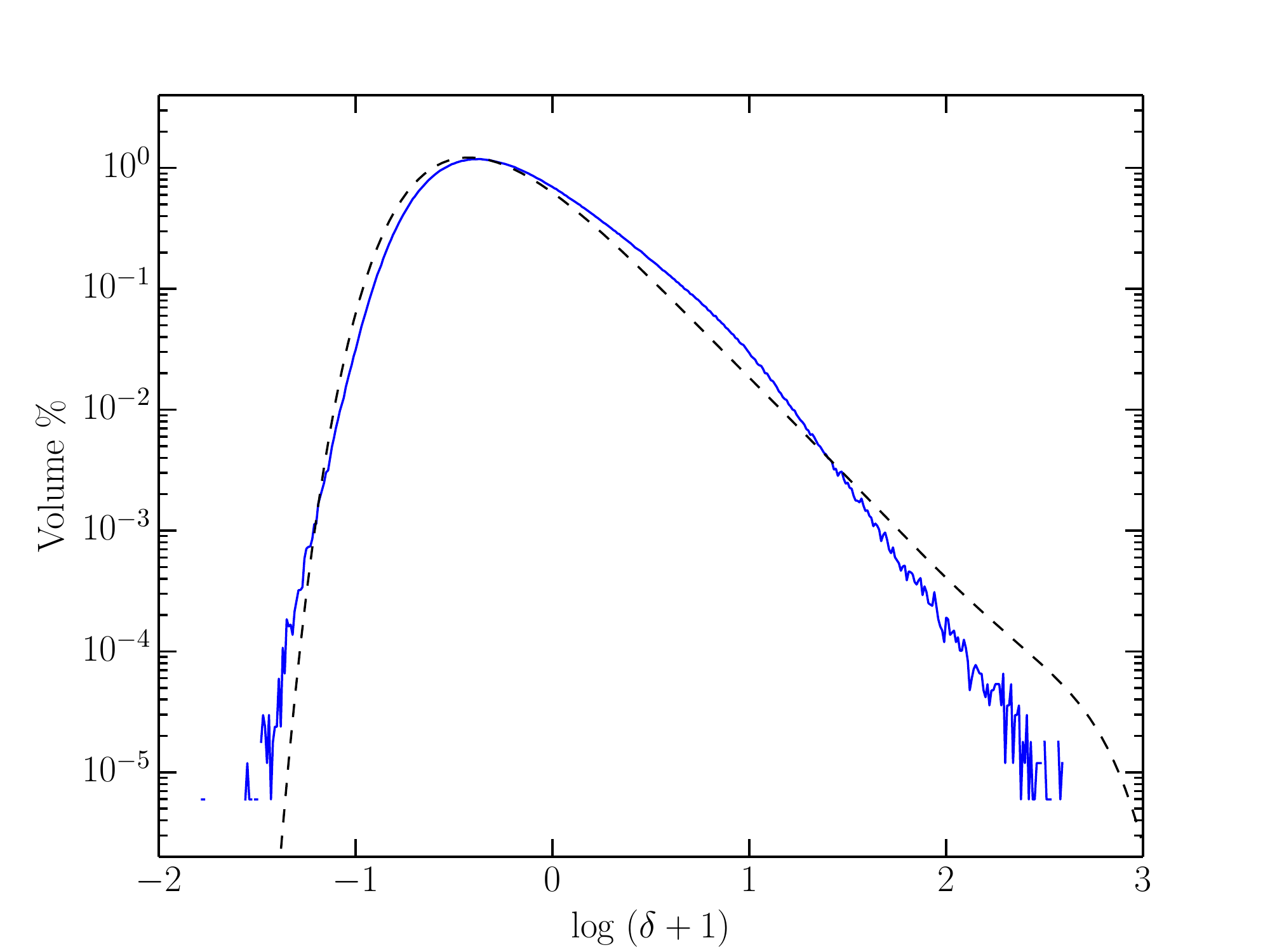}
\caption{Per centage of volume occupied by a given overdensity computed from a cosmological hydrodynamical simulation box of 60 h$^{-1}$ comoving Mpc at $z=2.8$ \citep[solid line,][]{viel13}. The overplotted curve (dashed line) is the result of a fit obtained from the higher resolution runs of \citet{boltonbecker}. See text for more details.per centages are given in bins of $\Delta [\log (1+\delta)] = 0.01$.}    
\label{fig:delta}
\end{center}
\end{figure}

In order to have a more physical insight on the volume filling factors
of the metal enriched gas, we analysed the  
output of a hydrodynamic simulation run with the {\sc gadget}-III
code \citep{springel05}, presented among the 
reference set of simulations in \citet{viel13}. This simulation has
reasonably well converged flux statistics for the \Lya\  forest and
evolve a standard concordance cosmological model with $2\times 512^3$
gas and dark matter particles in a periodic box of linear size
$60\ h^{-1}$ comoving Mpc, while 
the IGM thermal history is in agreement with observational data.
The gas particle distribution extracted at $z=2.8$ is then
interpolated on a grid using a Counts in Cells  scheme that allows us to
obtain an overdensity value over a grid of $512^3$ points.  
 

The overdensity values are reported in Fig.~\ref{fig:delta} in the
form of per centage of volume occupied in bins of $\Delta [\log
  (1+\delta)] = 0.01$ (solid line).  
However, since the reference simulation that we used is likely to be
not perfectly converged in terms of density probability distribution
function, we decided to overplot also the results obtained from a
higher resolution study by \citet[][dashed curve]{boltonbecker}. For
the densities of interest here, there is a small but not negligible
difference between the two curves and quoted values conservatively
bracket the uncertainties between the two results. 
 
\begin{table}
\caption{Volume filling factors of enriched gas in per centage derived from the two considered simulations: $^a$\citet{viel13} and $^b$\citet{boltonbecker}. The overdensity range corresponding to the inspected column densities is computed at the redshift of the simulations, $z=2.8$. We define $M \equiv \log Z/Z_{\sun}$. } 
\begin{center}
\label{tab:volume}
\begin{tabular}{r c  c  c}
\hline
$\log N_{\rm HI}$ & $[13.5:14.0]$  & $[14.0:14.8]$   & $[13.5:14.8]$ \\
$(\delta +1)$ & $[1.3:3.1]$ & $[3.1:11.8]$  & $[1.3:11.8]$ \\
\hline 
\multicolumn{4}{c}{ Reference model$^a$} \\
\hline 
Tot. Vol. &  $\sim12.7$ &  $\sim4.7$ & $\sim17.4$  \\
$M\gsim-3$ & $\sim7.6$ & $\sim2.8$ & $\sim10.4$ \\
$M < -3$ & $\sim5.1$ & $\sim1.9$ & $\sim7.0$\\
 \hline
 \multicolumn{4}{c}{ High-resolution model$^b$} \\
 \hline 
Tot. Vol. &  $\sim10.1$ &  $\sim3.0$  & $\sim13.1$\\
$M\gsim-3$ & $\sim6.1$ & $\sim1.8$ & $\sim7.9$\\
$M < -3$ & $\sim4.0$ & $\sim1.2$ & $\sim5.2$\\
 \hline
\end{tabular}
\end{center}
\end{table}

In the reference  simulation box, the overdensities $3.1 \le (\delta +
1) < 11.8$ (corresponding to the column density range $14.0 \le \log
N_{\rm HI} < 14.8$ at $z=2.8$) occupy  $\sim 4.7$ 
per cent of the volume. Using the per centage of absorbers observed at
different metallicities, we estimate that  $\sim 2.8$ per cent of 
the volume is enriched at $\log Z/Z_{\sun}\gsim-3$ and $\sim 1.9$
per cent of the volume has a metallicity $\log Z/Z_{\sun} < -3$.  
If we  extend our analysis to the mean density (corresponding in
Fig.~\ref{fig:CIV_OVI} to $\log N_{\rm HI} \sim 
13.5$) then, the probed volume would increase to $\sim 17.4$
per cent. Assuming the same per centages of absorbers at the different
metallicites hold to this lower densities, the volume enriched above a
metallicity  $\log Z/Z_{\sun}\sim-3$ becomes 
$\sim 10.4$ per cent.    
A conservative upper limit to the enriched volume  is determined from
our limits on \OVI: a maximum of 76 per cent of the gas with
overdensities between  $(\delta + 1) \simeq 1.3$ and $3.1$
(corresponding to the column density range $13.5 \le \log N_{\rm HI} <
14.0$ at $z=2.8$) is enriched to $\log Z/Z_{\sun}\gsim-3$. This
corresponds to an enriched volume fraction of $\sim9.6$ per cent which gives
an upper limit of $\sim12.5$ per cent adding the contribution of the larger
overdensities.  

The higher resolution simulation predicts a smaller volume occupation
for the range of overdensities considered in this work.  In
particular, for the column density range $14.0 \le \log N_{\rm HI} <
14.8$ the occupied volume at $z=2.8$ is 3.0 per cent, for the range
$13.5 \le \log N_{\rm HI} < 14.0$ is $\sim10.1$ per cent. These numbers
translate into a maximum volume enriched to $\log Z/Z_{\sun}\gsim-3$
of $\sim 9.5$ per cent.  
All the volume filling factors are detailed in Table~\ref{tab:volume}. 

It is relevant to note that the contribution of the regions traced by
\HI\ column densities $\log N_{\rm HI} > 14.8$ is negligible in terms
of enriched volume as can be seen in Fig.~\ref{fig:delta}.  

\subsection{Dependence on redshift}
\begin{figure*}
\begin{center}
\includegraphics[width=18cm]{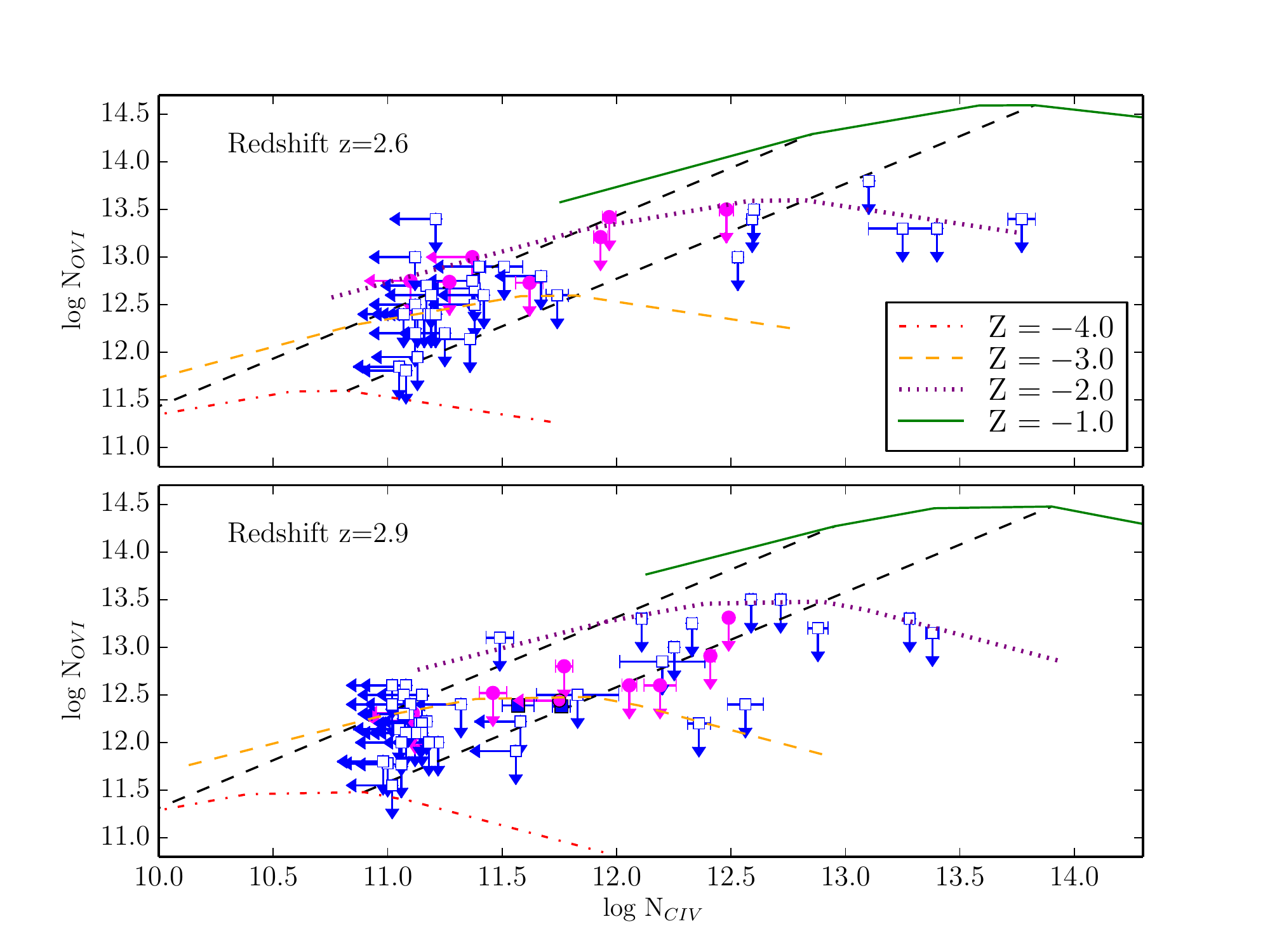}
\caption{Absorption systems in our sample for which we have either an
  upper limit or a detection for \CIV\ and \OVI\ divided into
  two redshift bins: $2.4902 \lsim z < 2.758$ (upper panel) and $2.758 \lsim z < 3.0255$ (lower
    panel). Systems for which one or both values are upper limits are
  plotted as  open symbols. Filled dots mark those systems for which \CIV\ is a detection and \OVI\ is a possible
  detection (only one of the line of the doublet is detected), while filled squares mark systems for which both \CIV\ and
  \OVI\ are detections (which for the \OVI\ means that both lines of the doublet were detected). 
  The superposed lines in the two plots are the predictions of the \cloudy\ models described in
  Section 5 for the four metallicities listed in the caption, computed at the average redshift of the samples, $z=2.6$
  (upper panel) and $z=2.9$ (lower panel). The black dashed lines join the model
  predictions for the four metallicities at $(\delta + 1)=3.7$ and 14.1 at
  $z=2.6$, and at  $(\delta + 1)=2.8$ and 10.8 at $z=2.9$, corresponding
  to \HI\ column densities of $\log N_{\rm HI} =14$ and 14.8,
  respectively.}    
\label{fig:CIV_OVI_z}
\end{center}
\end{figure*}

The association of \HI\ lines with metal lines is a fundamental step
which allows us to estimate the metallicity of the gas at different
overdensities which, however, is always affected by a certain degree
of uncertainty. If, for example, the metals are transported in the IGM
by small metal-rich clumps that then lose their identity in \HI, as
suggested by \citet[][]{schaye07}, many of the
metallicities we are deriving could be considered only as average
values and the real picture would be much more complicated.  

With the aim of verifying if the results plotted in
Figs.~\ref{fig:CIV_OVI} and \ref{fig:metal_hist} are reliable, we have considered only those
absorbers for which we had either a detection or an upper limit for
both \CIV\ and \OVI, 
disregarding the information on \HI.  
Then, we have also investigated the possible effects of redshift,
dividing the obtained sample in two redshift bins: $2.4902 \lsim z <
2.758$ and $2.758 \lsim z < 3.0255$.    

The results are shown in Fig.~\ref{fig:CIV_OVI_z}, where the two
redshift subsamples are compared with the predictions of the \cloudy\
models described before, computed at the average redshifts of the
sub-samples, $z=2.6$ and $z=2.9$,
respectively. The black, dashed lines mark the regions between 
$\log N_{\rm HI} =14$ and 14.8 (see Fig.~\ref{fig:CIV_OVI}), and the column
densities have been translated into overdensity values as already
described in the previous section.  

There is no evidence of a significant difference in the metallicity of
the investigated gas in the two considered redshift bins.  

We confirm that in the overdensity regime where our observations are
most sensitive, from $(\delta +1) \simeq 2.8-3.7$ to $(\delta +1) \simeq 10.8-14.1$, most of our
measures indicate a metallicity $-3 \lsim Z \lsim-2$. In particular,
the only two systems for which we have detected both \CIV\ and
\OVI\ are at an overdensity of $\sim 5-10$ and at a metallicity
$Z\sim-3$.  
Considering the total sample (88 absorbers), we find that $\sim 40$
per cent of the systems have a metallicity  $Z<-3$.  
This per centage should be considered as a lower limit and would
likely increase with the increase of the \OVI\ statistics and of the
SNR of the spectra.  

\section{Conclusions}
In this work, we presented the analysis of the ultra-high SNR, high
resolution  spectrum (dubbed the {\it deep} spectrum) of the quasar HE0940-1050 ($z_{\rm em}=3.0932$) obtained with the UVES spectrograph at the ESO VLT telescope. 
The aim of this observation was to shed light on the abundance and 
distribution of metals in the low density intergalactic gas at
$z\sim3$ in order to investigate the details of the enrichment
process.   

 The investigated redshift range, $2.4902 \lsim z \lsim 3.0255$, was chosen in order to have at least the transitions \Lya\ and \Lyb\ of the \HI\ Lyman series to allow a reliable determination of $N_{\rm HI}$.
In this whole redshift interval the deep spectrum is sensitive at $3\ \sigma$ to \CIV\ $\lambda\,1548$ \AA\ lines with column densities down to $\log N_{\rm CIV} \simeq 11.4$
(assuming a Doppler value $b=7$ \kms), in 90 per cent of this interval the spectrum is sensitive to \CIV\ lines down to $\log N_{\rm CIV} \simeq 11.3$ and in 60 per cent down to $\simeq11.1$. 

We searched also for \OVI\ in same redshift interval. 
The search for weak \OVI\ lines at $z\sim3$ is hindered by the crowded
\Lya\ and \Lyb\ forests. Indeed, we could detect (i.e. both lines of
doublet were observed) only three weak \OVI\ absorbers, of which two
were identified independently of their association with a
\CIV\ absorber. 

The observational data were then compared with the results of a grid of \cloudy\ models obtained assuming solar chemical composition and varying the metallicity, the overdensity and the temperature of the slab of gas (see Section~5). 



The main conclusions that can be drawn from our study are the following. 

\noindent
 1. All \HI\ lines in our sample with column density $\log N_{\rm HI}
\ge 14.8$ show an associated \CIV\ absorption. In the range $14.0 \le
\log N_{\rm HI} < 14.8$, 43 per cent of \HI\ lines has an associated
\CIV\ absorption with column density down to our sensitivity limit. 
At column densities $\log N_{\rm HI} < 14.0$, the detection rates drop to less than 10 per cent. 
The comparison with the outcome of our grid of \cloudy\ models shows that even if the gas traced by these
\HI\ column densities was as enriched as the one at larger column
densities, our spectrum would not have the sensitivity to probe it.  

\noindent
2. The fraction of \HI\ lines with $\log N_{\rm HI} \ge 14$ showing
associated \CIV\  detections in our sample is at least a factor of 2 larger than the
fraction of \HI\ lines in the same column density range, tracing the CGM of relatively bright Lyman
break galaxies hosted by dark matter haloes of average mass $\sim
10^{12}$ M$_{\sun}$, as found by \citet[][see
  Fig.~\ref{fig:CIV_detrate}]{rudie12}.  This result suggests that metals could lie also at larger distances than 300 physical kpc from the considered galaxies but also that they could be present around lower mass galaxies,  in agreement with a variety of theoretical studies
\citep[e.g.][]{madau01,scannapieco02,samui08,oppenheimer09,booth12}.  
 
\noindent 
3. In the range of overdensities in which our \CIV\ data are most
sensitive to low metallicities, from 
$(\delta+1)\simeq 3.1$ to $(\delta+1)\simeq11.8$ (corresponding to the
range of column densities $14 \le \log N_{\rm HI} < 14.8$ at $z=2.8$),  the comparison with our grid of \cloudy\ models indicates that between $\sim38$ and 60 per cent of the absorbers are enriched to metallicities $-3 \lsim
\log Z/Z_{\sun} \lsim -1$, while the remaining $\sim40-62$ per cent have a
metallicity lower than $-3$   (see Figs.~\ref{fig:CIV_OVI} and
\ref{fig:metal_hist}). 
 
\noindent
4. Through a comparison with the predictions of two cosmological
simulations, we derived that the volume filling factor of the IGM  gas
(down to the mean density) enriched to a metallicity $\log Z/Z_{\sun} \gsim -3$ should be of the order of $\sim 10-13$
per cent.  This is again in agreement with predictions by theoretical
studies in which the IGM was enriched at high redshift by low mass
objects.  

As a final statement, we would like to underline that the main
weakness of this study is that it is based on a single high-resolution
quasar spectrum albeit with an exceedingly high signal-to-noise
ratio\footnote{The SNR per resolution element varies between $\sim200$ and 500 in the \CIV\ forest at $2.4902 \lsim z \lsim 3.0255$, being larger than $\sim350$ in more than 80 per cent of the redshift interval.}. Such amazing data are the result of a gigantic observational
effort performed with the UVES spectrograph at the VLT, which will
unlikely be repeatable on a short time-scale. As a
consequence it will be difficult to have soon a significative sample
of exceptional spectra as this one.

 In the near future, the ESPRESSO spectrograph \citep{pepe14}, a
  second generation VLT instrument whose first light is expected in
  2017, will provide an opportunity for further high SNR studies of
  quasars. It will offer the unique possibility of working at a
  resolution $R\sim60,000$ collecting   the light from all four unit telescopes at
  the same time,  mimicking the outcome of a 16m-equivalent
  telescope.  
For the real breakthrough in this research field, however,  we will have to wait
10-15 years until the next generation of high-resolution spectrographs
working on 30m-class telescopes will be on stage.


\section*{Acknowledgements}

Based on observations collected at  the  European  Southern  Observatory  Very  Large  Telescope,  Cerro Paranal,  Chile -- Programs 166.A-0106, 079.B-0469, 185.A-0745, 092.A-0170. 
We are indebted to an anonymous referee for helpful comments that clarified several points in the paper. 
VD is grateful to Gwen Rudie for sharing and adapting her results on
the KBSS for Fig.~\ref{fig:CIV_detrate}.  
MV is supported by the ERC Starting Grant ``cosmoIGM'' and PD51 INDARK
grant. TSK acknowledges funding support from the ERC Starting Grant
``cosmoIGM'', through grant GA-257670. MH was supported by the ERC
Advanced Grant 320596 ``The Emergence of Structure during the epoch of
Reionisation''. 
PB is supported by the INAF PRIN-2014 grant ``Windy black holes combing galaxy evolution''. 
Parts of this research were conducted by the Australian Research
Council Centre of Excellence for All-sky Astrophysics (CAASTRO),
through project number CE110001020.

\appendix


\section{Plots of possible \OVI\ detections}

\begin{figure}
\begin{center}
\includegraphics[width=9.cm]{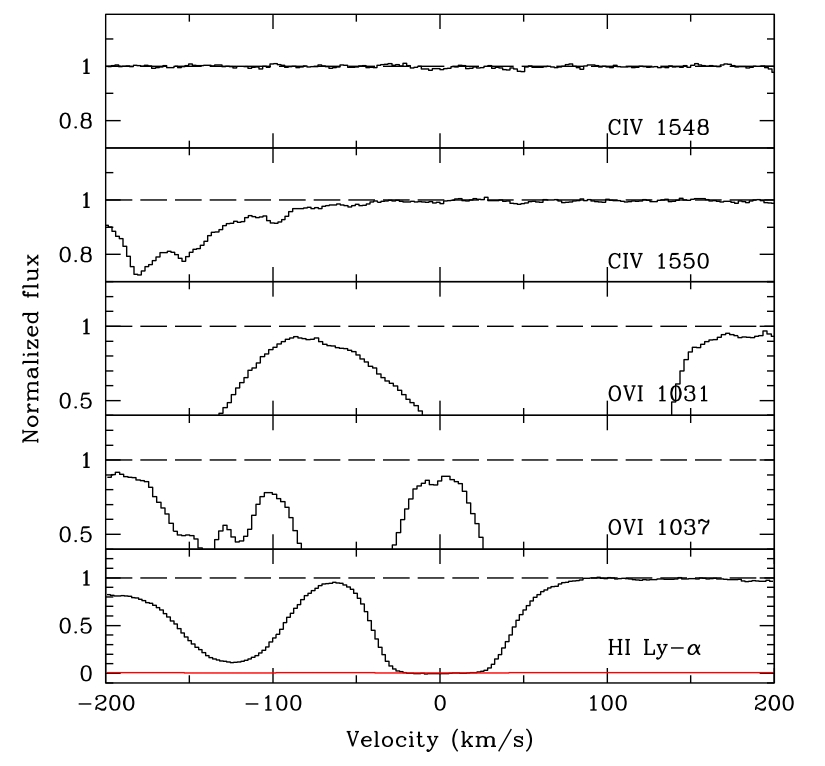}
\caption{Possible detection of the transition \OVI\ $\lambda\,1038$ at
$z_{\rm abs}=2.507588$ corresponding to $v=0$ \kms\ in the plot. \CIV\ is not detected at this redshift.}    
\label{fig1:z2p507}
\end{center}
\end{figure}

\begin{figure}
\begin{center}
\includegraphics[width=9.cm]{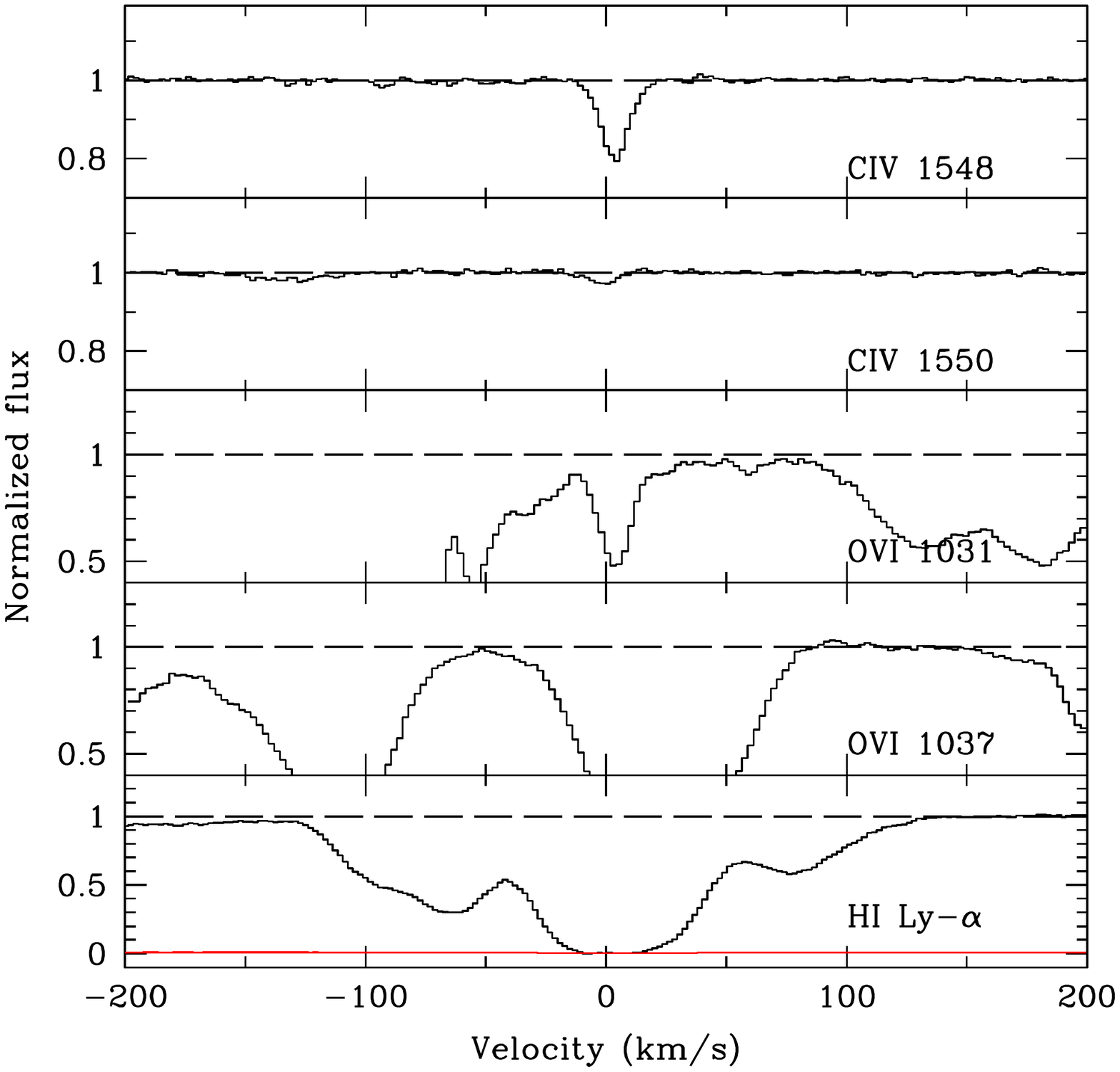}
\caption{Detection of the \CIV\ doublet (the transition at
  $\lambda\,1548$ is blended) and possible detection of the transition \OVI\ $\lambda\,1032$ at
$z_{\rm abs}=2.566402$.}    
\label{fig2:z2p56}
\end{center}
\end{figure}

\begin{figure}
\begin{center}
\includegraphics[width=9.cm]{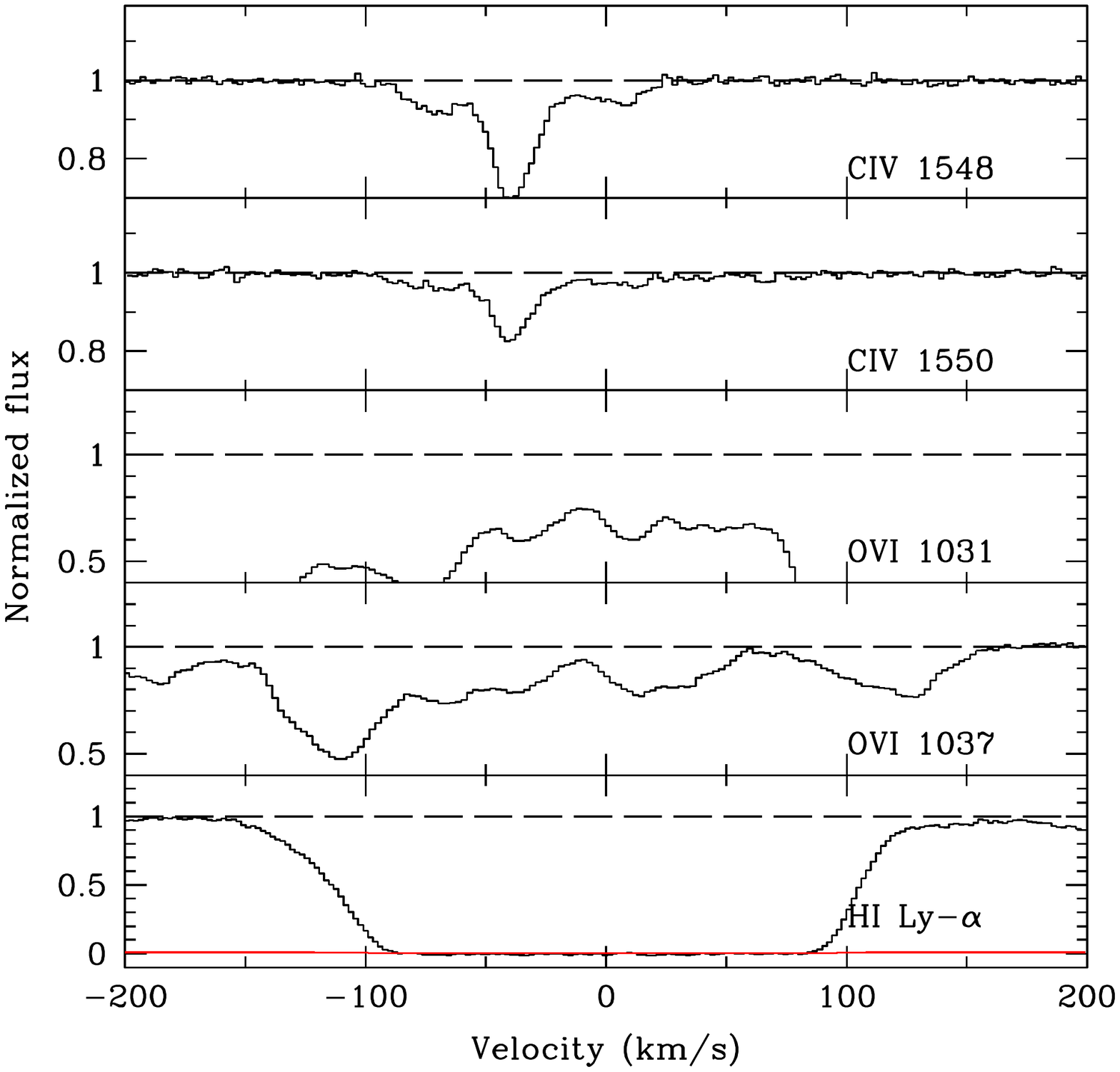}
\caption{Detection of the \CIV\ doublet and
  possible detection of the transition \OVI\ $\lambda\,1032$ at $z_{\rm abs}=2.643643$.}    
\label{fig3:z2p643}
\end{center}
\end{figure}

\begin{figure}
\begin{center}
\includegraphics[width=9.cm]{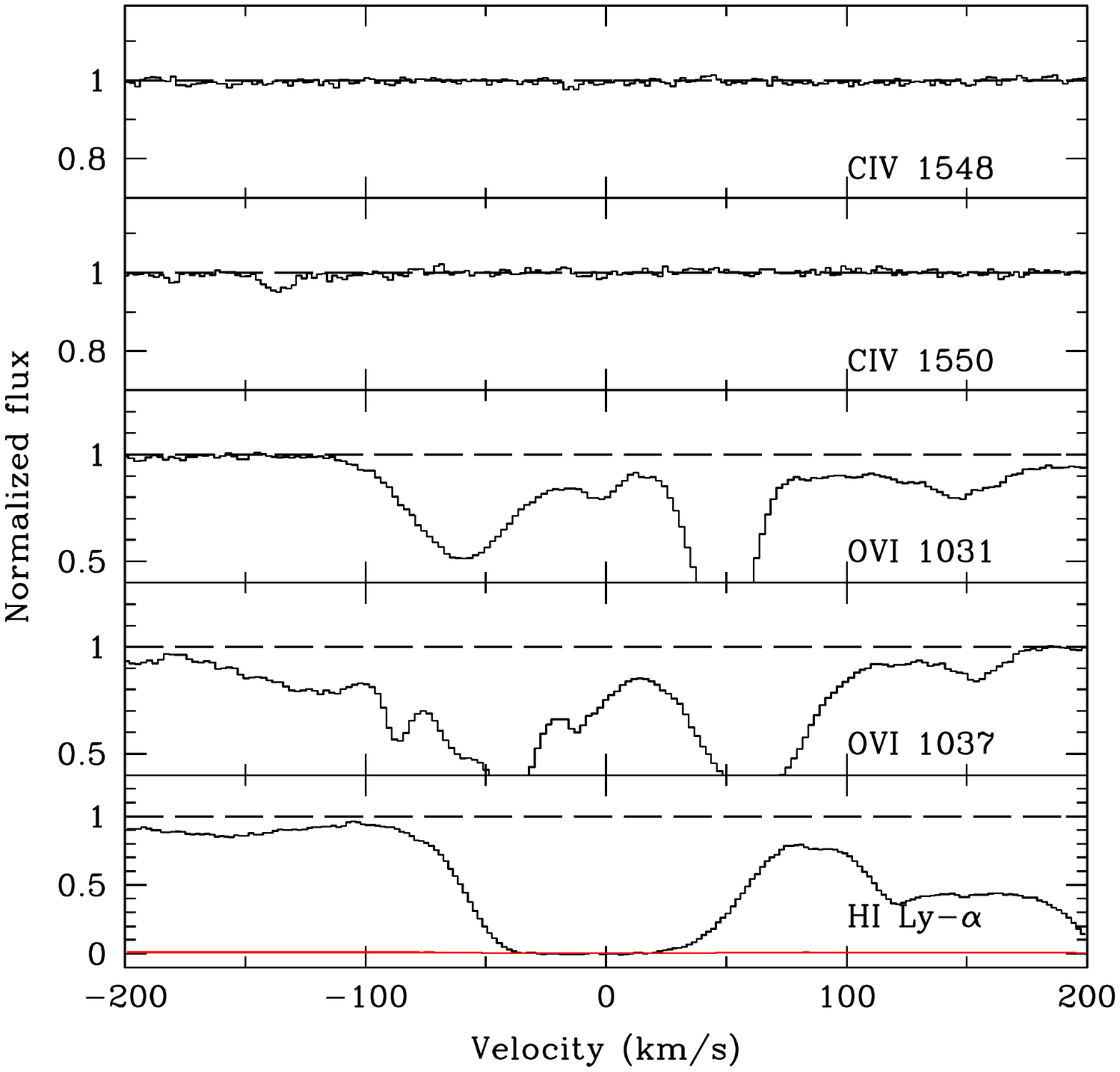}
\caption{Possible detection of the transition \OVI\ $\lambda\,1032$ at
$z_{\rm abs}=2.654839$. \CIV\ is not detected at this redshift.}    
\label{fig4:z2p654}
\end{center}
\end{figure}

\begin{figure}
\begin{center}
\includegraphics[width=9.cm]{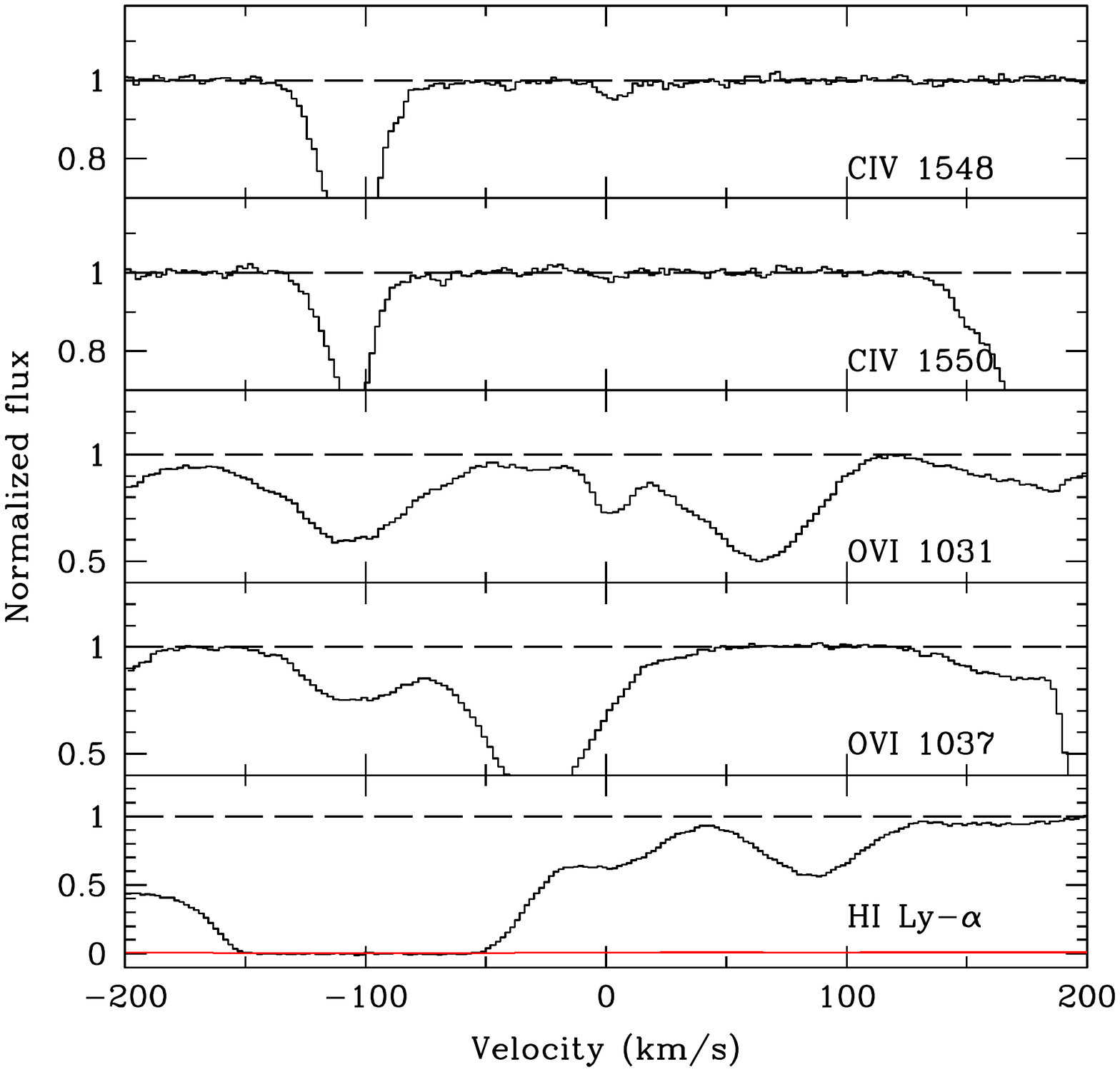}
\caption{Detection of the \CIV\ doublet and possible detection of the
  transition \OVI\ $\lambda\,1032$ at $z_{\rm abs}=2.659215$.}    
\label{fig5:z2p659}
\end{center}
\end{figure}

\begin{figure}
\begin{center}
\includegraphics[width=9.cm]{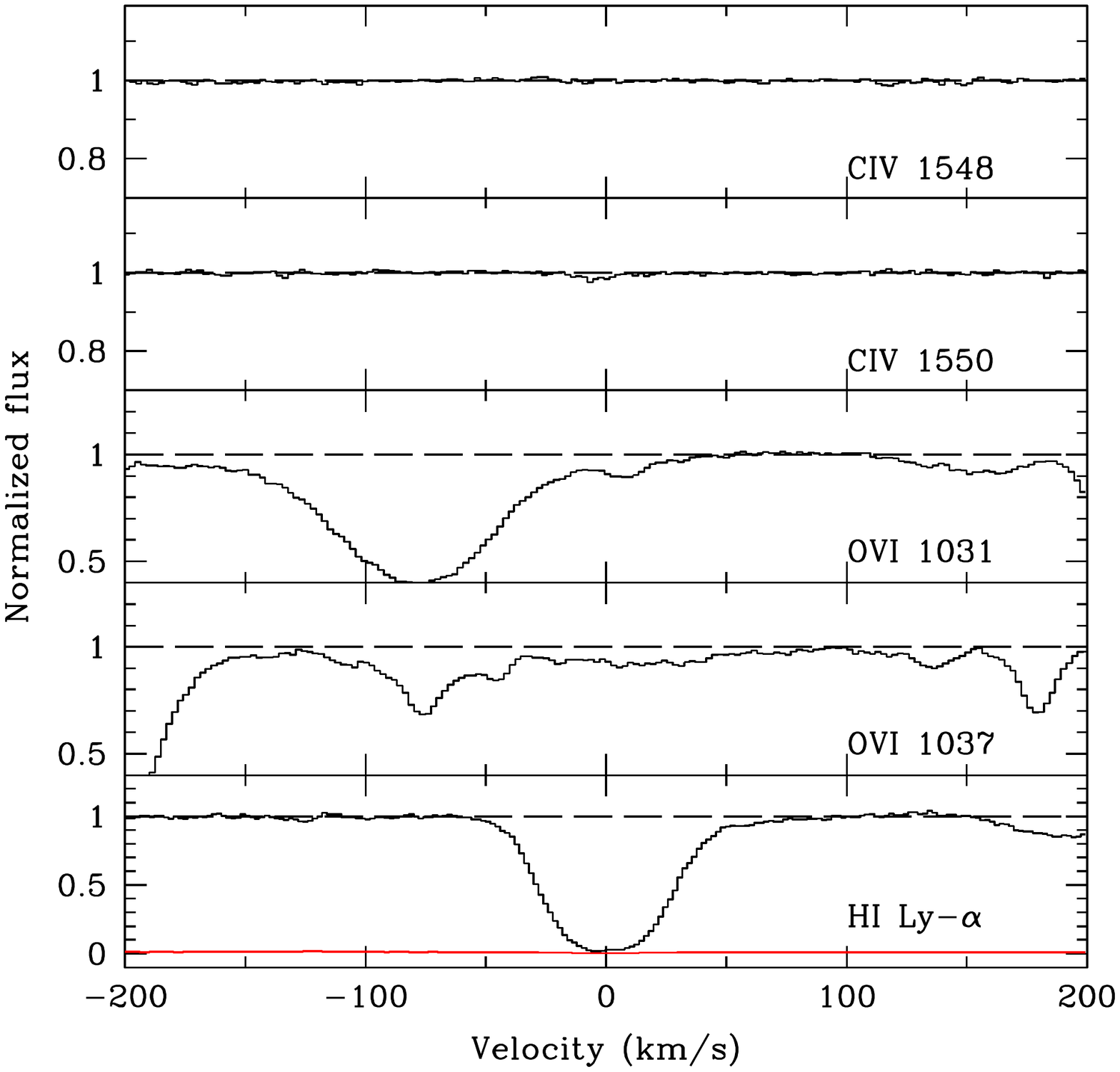}
\caption{Possible detection of the transition \OVI\ $\lambda\,1032$ at
$z_{\rm abs}=2.705568$. \CIV\ is not detected at this redshift. }    
\label{fig6:z2p70}
\end{center}
\end{figure}

\begin{figure}
\begin{center}
\includegraphics[width=9.cm]{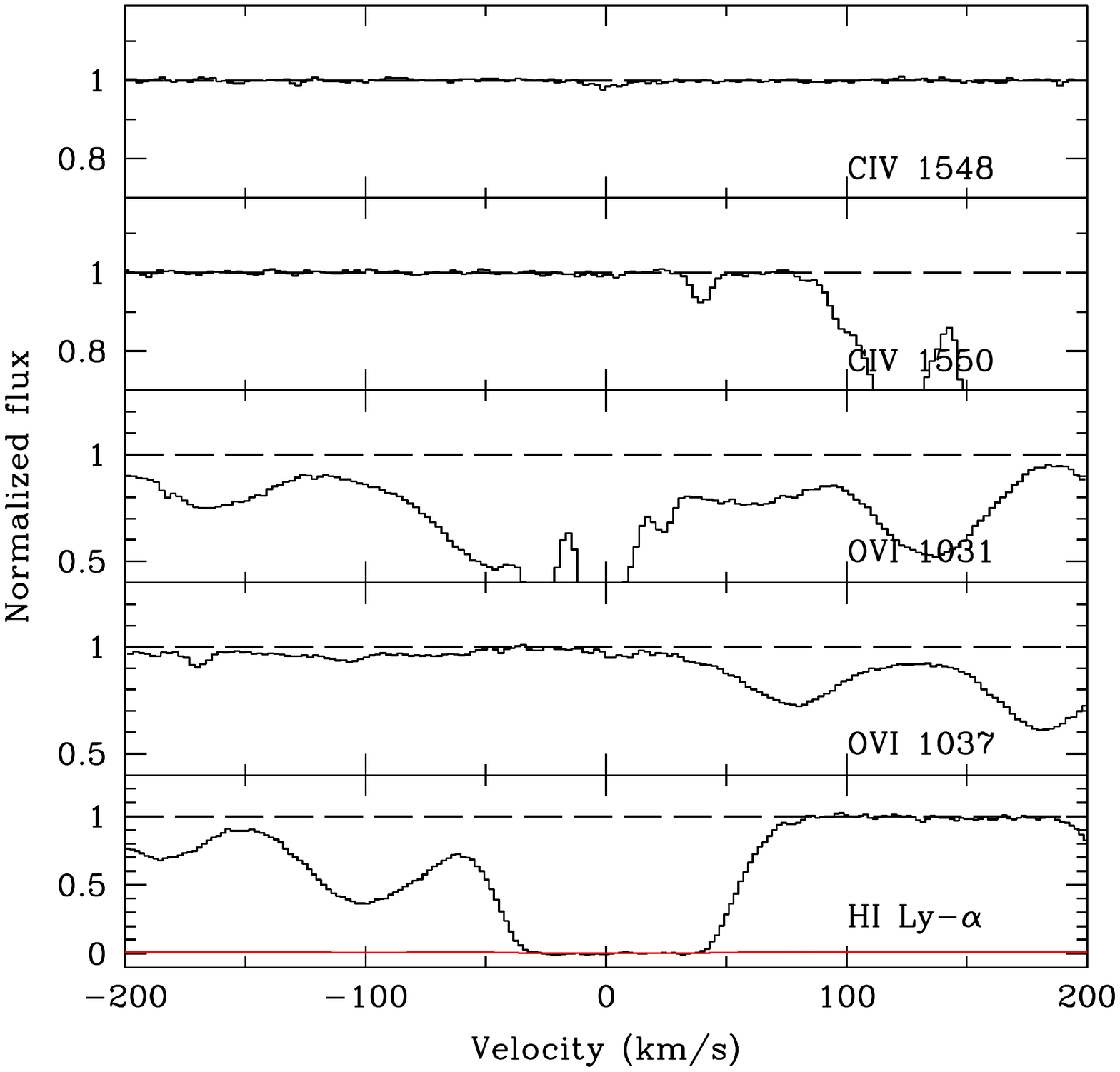}
\caption{Detection of the weak \CIV\ transition $\lambda\,1548$ and
  possible detection of the transition \OVI\ $\lambda\,1032$ at $z_{\rm abs}=2.71167$.}    
\label{fig7:z2p711}
\end{center}
\end{figure}

\begin{figure}
\begin{center}
\includegraphics[width=9.cm]{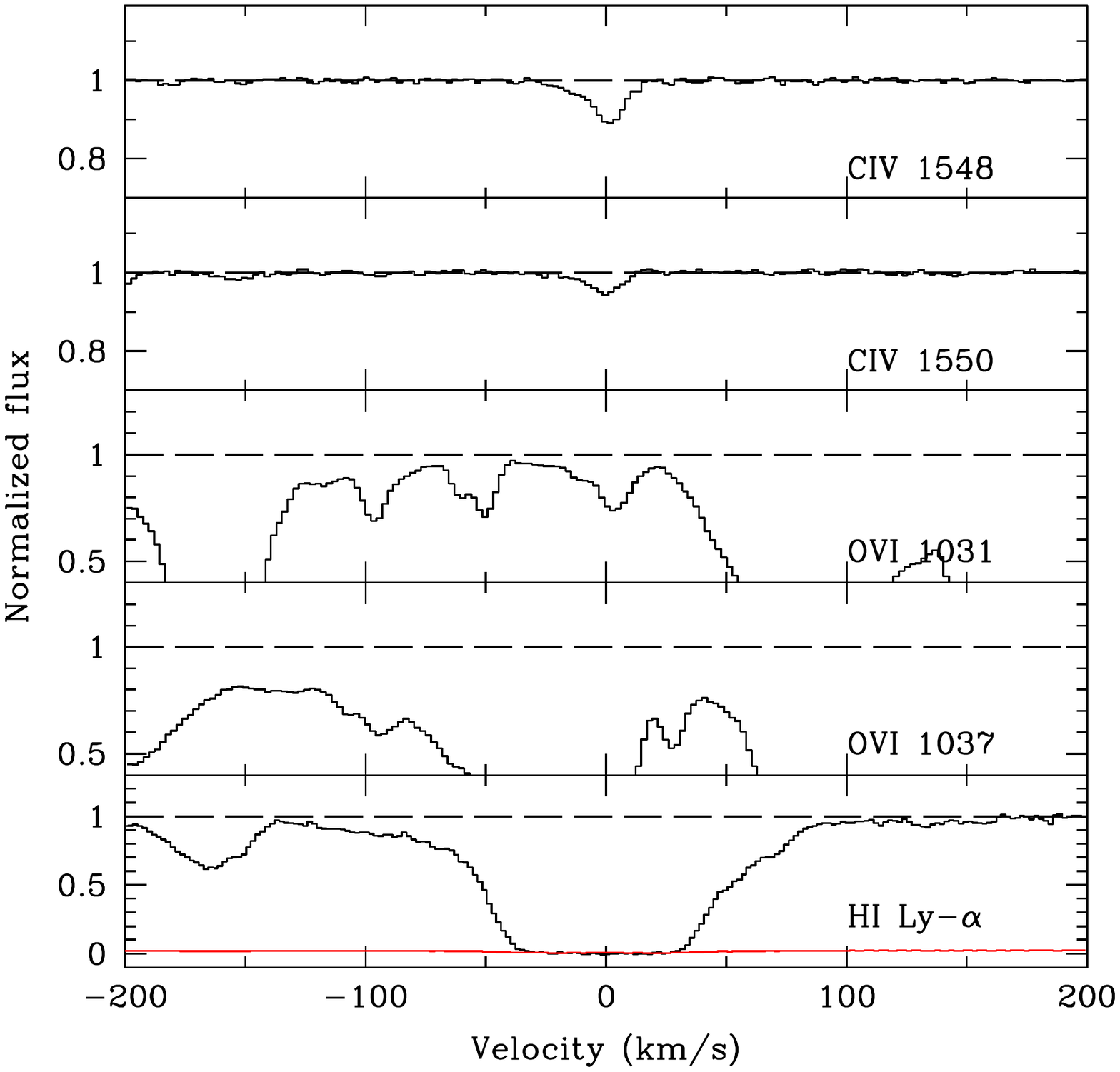}
\caption{Detection of the \CIV\ doublet and possible detection of the
  transition \OVI\ $\lambda\,1032$ at $z_{\rm abs}=2.771388$.}    
\label{fig8:z2p77}
\end{center}
\end{figure}

\begin{figure}
\begin{center}
\includegraphics[width=9.cm]{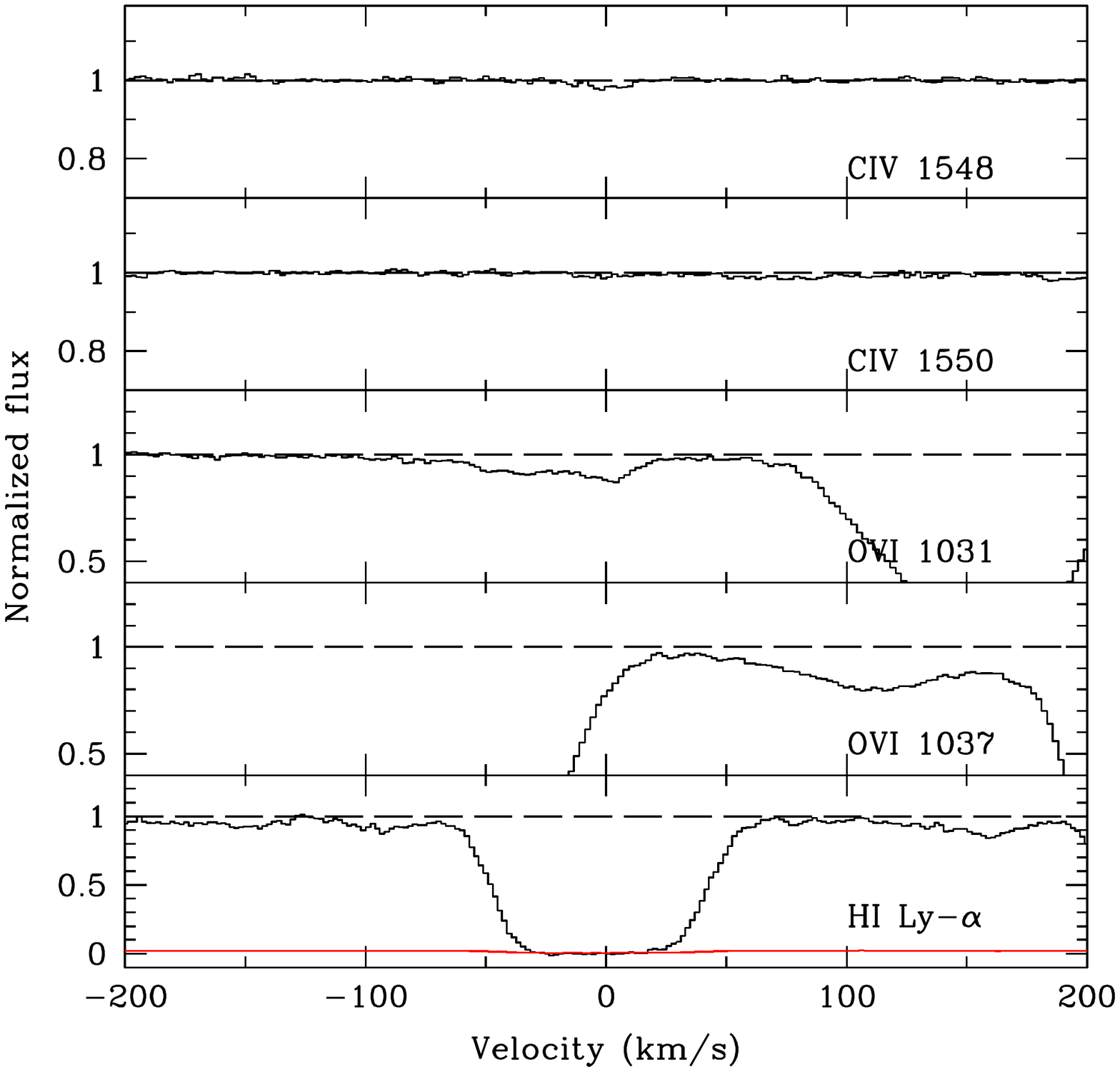}
\caption{Detection of the weak \CIV\ transition $\lambda\,1548$ and
  possible detection of the transition \OVI\ $\lambda\,1032$ at $z_{\rm abs}=2.785936$.}    
\label{fig9:z2p78}
\end{center}
\end{figure}

\begin{figure}
\begin{center}
\includegraphics[width=9.cm]{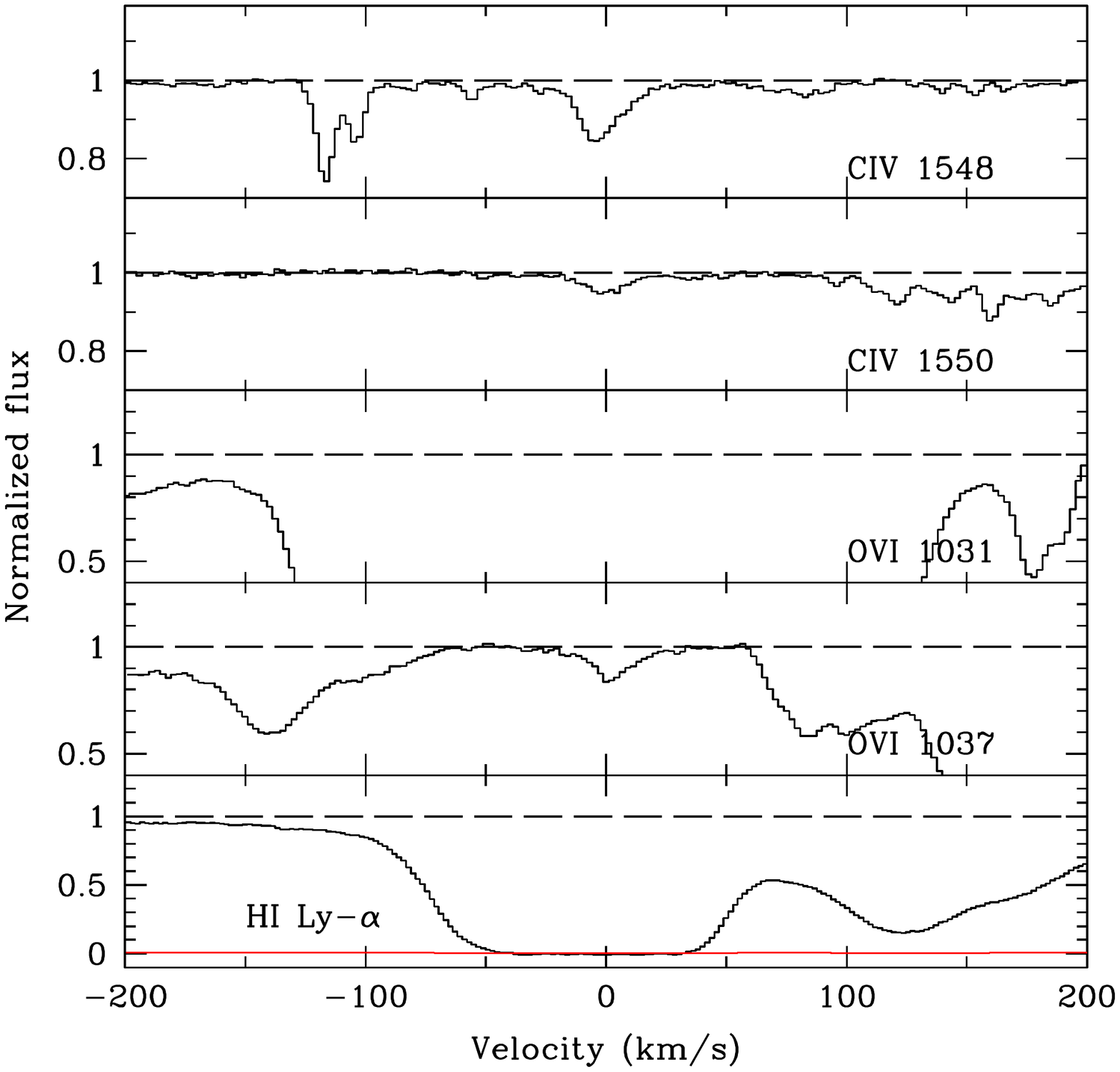}
\caption{Detection of the \CIV\ doublet and possible detection of the
  transition \OVI\ $\lambda\,1038$ at $z_{\rm abs}=2.810878$.}    
\label{fig10:z2p81}
\end{center}
\end{figure}

\begin{figure}
\begin{center}
\includegraphics[width=9.cm]{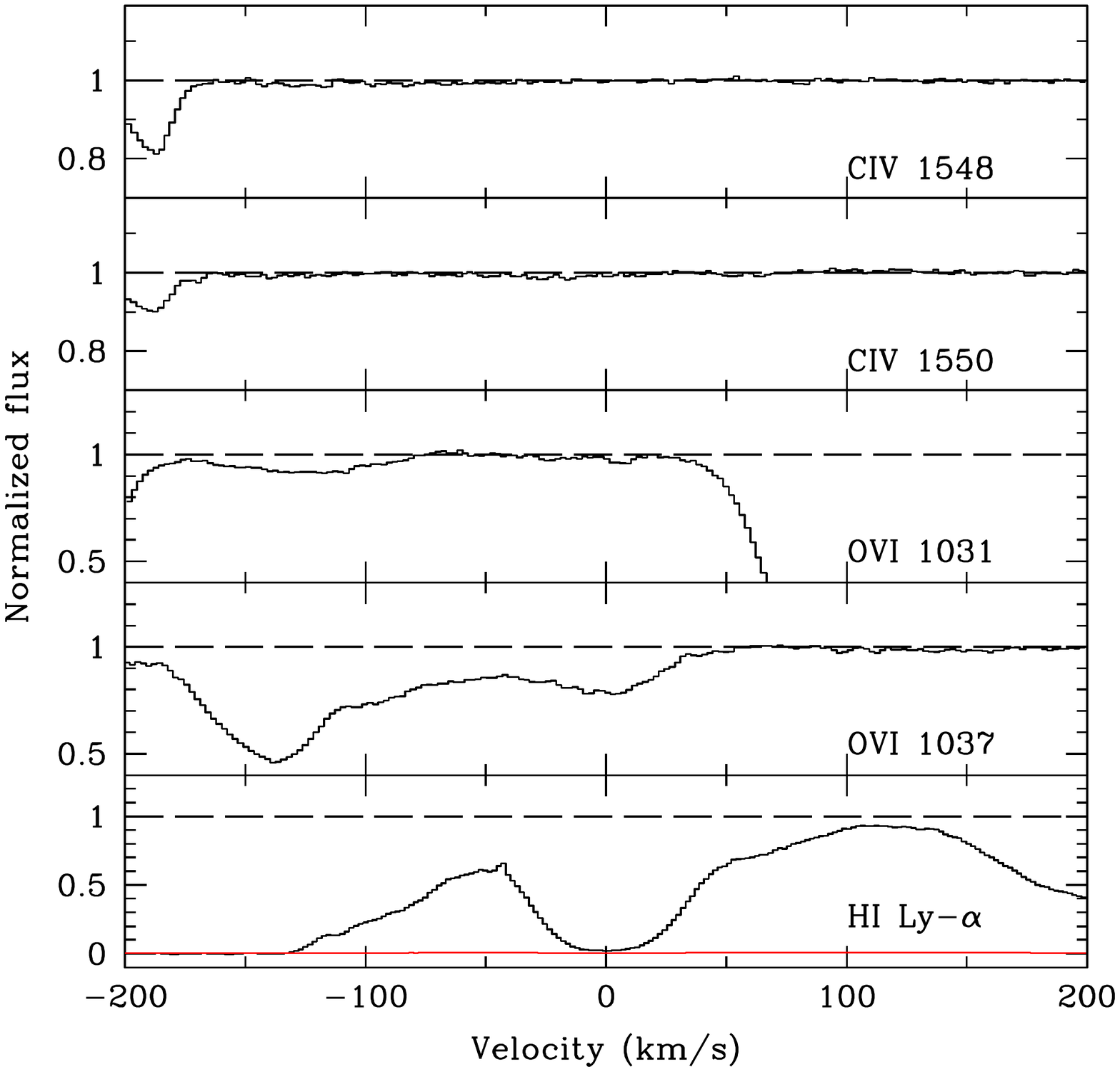}
\caption{Possible detection of the transition \OVI\ $\lambda\,1032$ at
  $z_{\rm abs}=2.863317$. \CIV\ is not detected at this redshift.}    
\label{fig11:z2p86}
\end{center}
\end{figure}

\begin{figure}
\begin{center}
\includegraphics[width=9.cm]{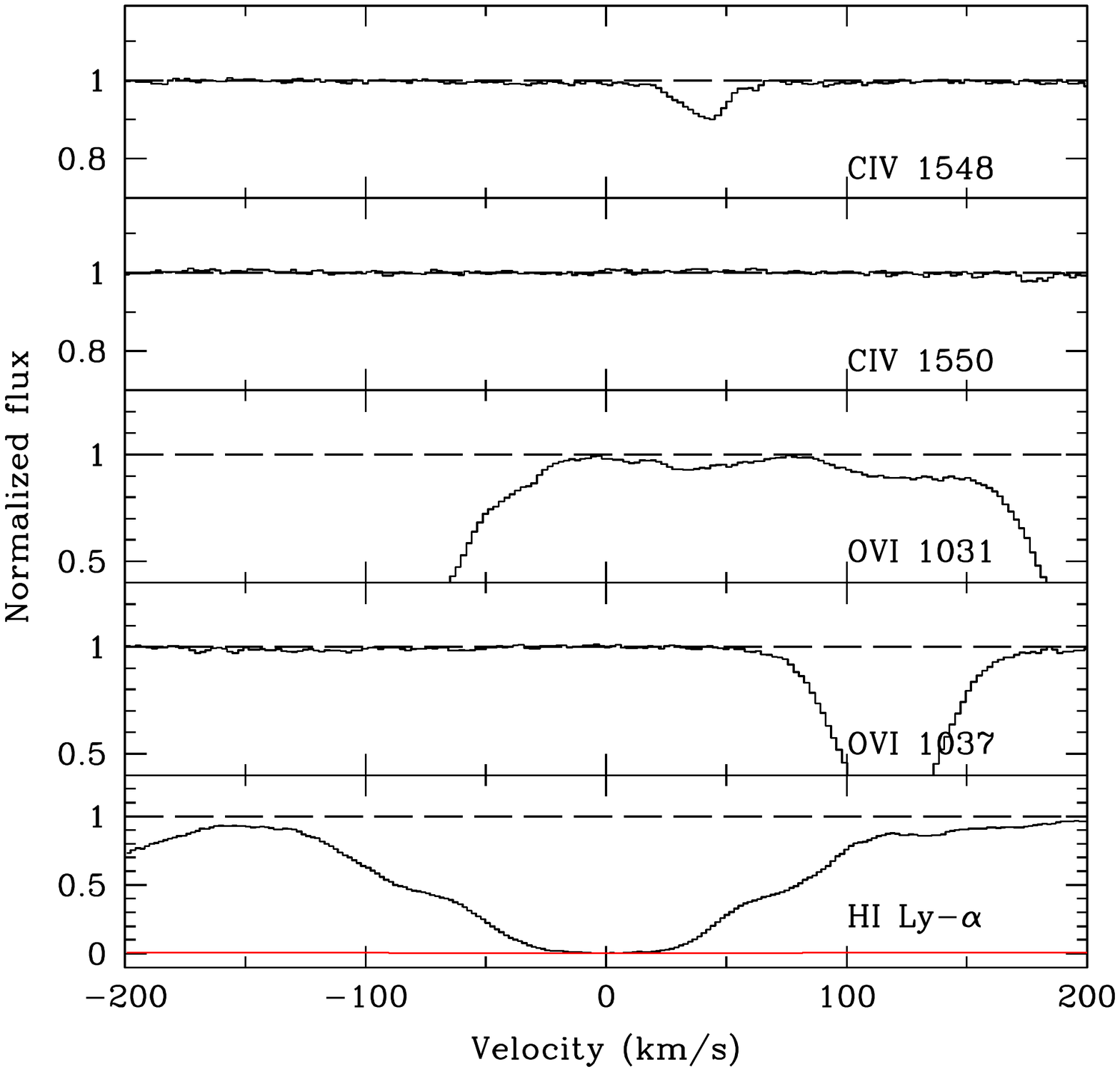}
\caption{Possible detection of the transition \OVI\ $\lambda\,1032$ at
$z_{\rm abs}=2.566402$. \CIV\ is not detected at this redshift. }    
\label{fig12:z2p866}
\end{center}
\end{figure}

\begin{figure}
\begin{center}
\includegraphics[width=9.cm]{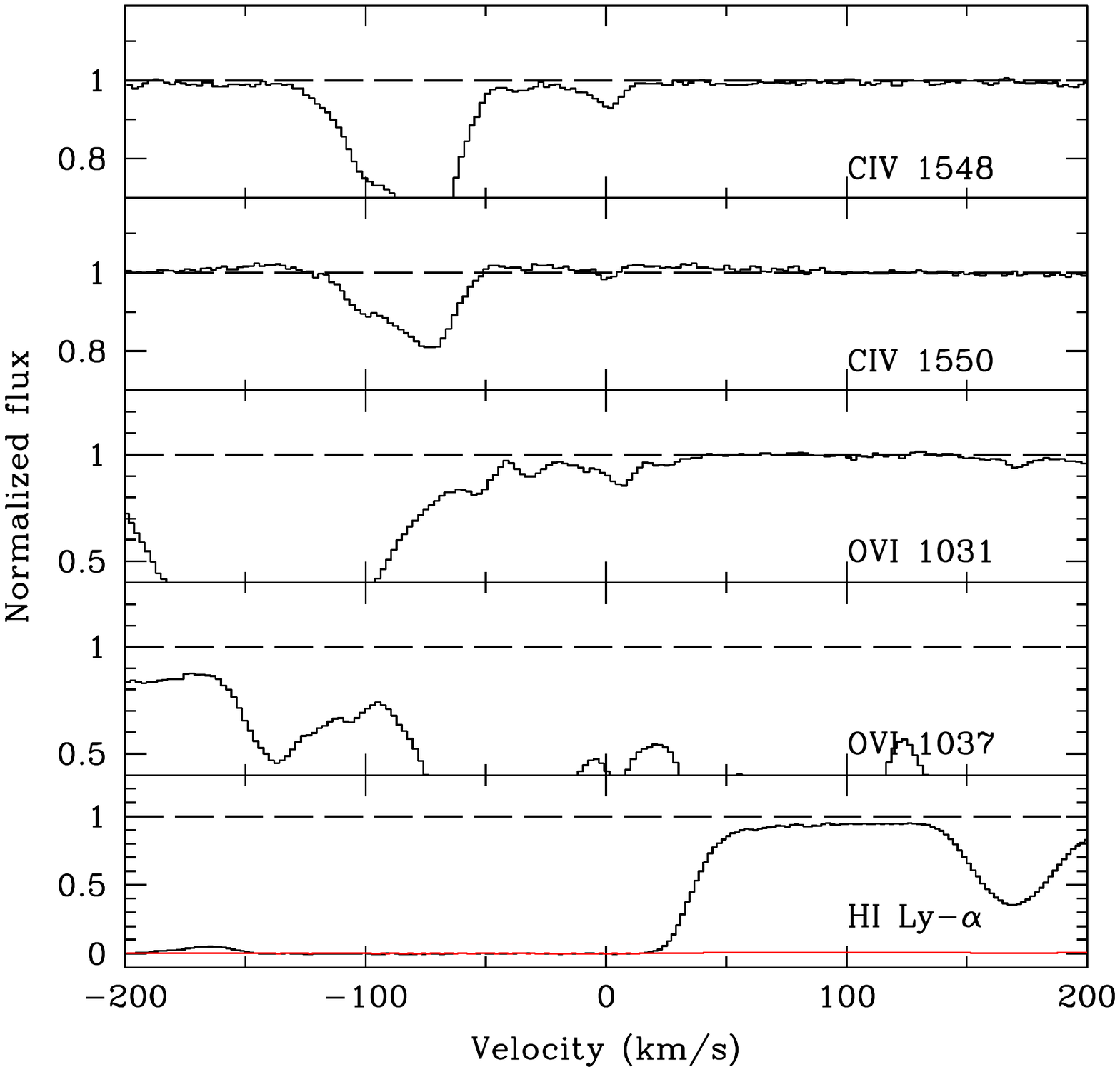}
\caption{Detection of the \CIV\ doublet  and possible detection of the
  transition \OVI\ $\lambda\,1032$ at $z_{\rm abs}=2.918023$.}    
\label{fig13:z2p918}
\end{center}
\end{figure}

\begin{figure}
\begin{center}
\includegraphics[width=9.cm]{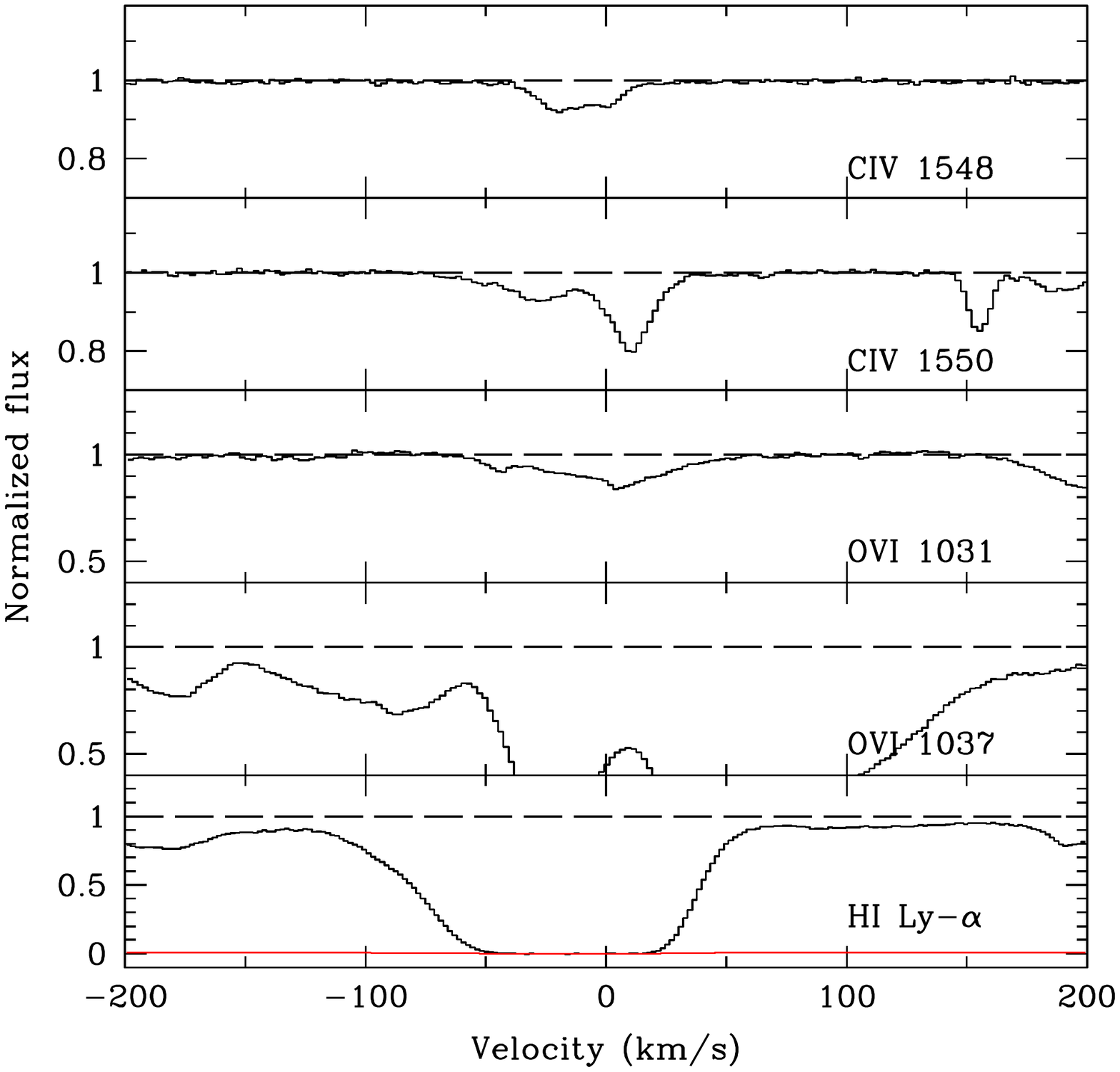}
\caption{Detection of the \CIV\ doublet (the transition
  $\lambda\,1551$ is blended) and possible detection of the transition
  \OVI\ $\lambda\,1032$ at $z_{\rm abs}=2.931062$.}    
\label{fig14:z2p931}
\end{center}
\end{figure}

\begin{figure}
\begin{center}
\includegraphics[width=9.cm]{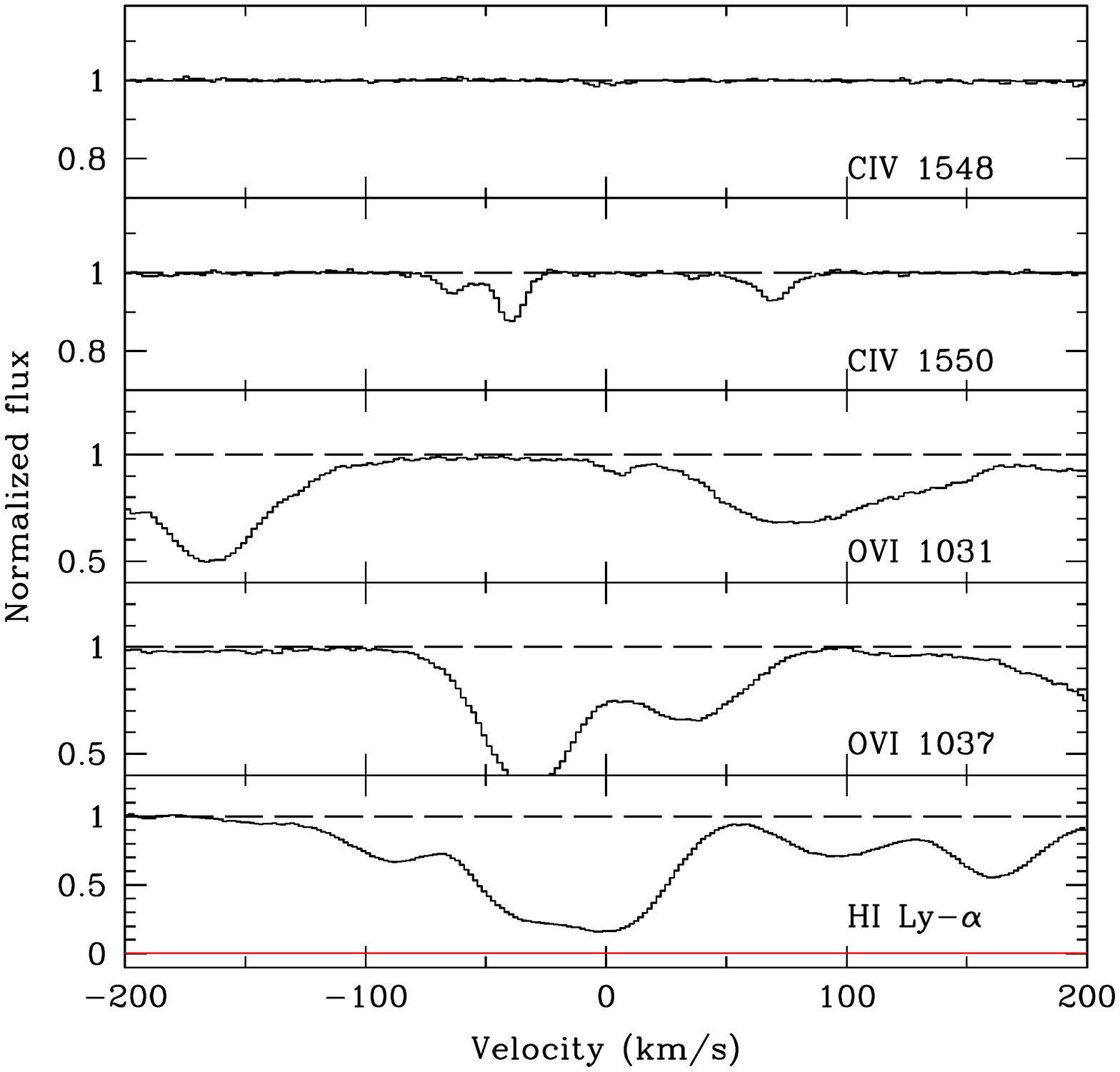}
\caption{Detection of the weak \CIV\ transition $\lambda\,1548$ and
  possible detection of the transition \OVI\ $\lambda\,1032$ at $z_{\rm abs}=2.98251$.}    
\label{fig15:z2p982}
\end{center}
\end{figure}

\label{lastpage}
\end{document}